 \newcommand{\id}{  \mathbb{1} }
 \newcommand{\zero}{  \mathbb{0} }
\newcommand{\Real}{\mathbb{R}\mathrm{e}}
\newcommand{\Imag}{\mathbb{I}\mathrm{m}}
\newcommand{\llangle}{\left\langle}
\newcommand{\rrangle}{\right\rangle}
\newcommand{\Tr}{\mathrm{Tr}}

\newcommand{\half}{\frac{1}{2}}

\newcommand{\UB}{ \mathbf{U}_{\textsc{B}} }
\newcommand{\SB}  { \mathbf{S}_{\textsc{B}} }
\newcommand{\HEX}{NPC}	
\newcommand{\vect}[1]{\boldsymbol{#1}}

\newcommand{\vQ}{ \hat{\vect{Q}}}                            
\newcommand{\Sy} {\vect{\Omega}}               
\newcommand{\V}{\vect{V}}                              
\newcommand{\PT}{\vect{\Gamma}}                 
\newcommand{\vd}{ {\mathbf{d}}}

\newcommand{ \PTA} [1]  { \mathsf{\, \scriptscriptstyle PT(#1 )}}    
\newcommand{\X} {\hat{X}}
\newcommand{\Y} {\hat{Y}}
\newcommand{\Q} {\hat{Q}}
\newcommand{\dQ} {\delta\hat{Q}}

\newcommand{\azero} {\hat{a}_{0}}
\newcommand{\auno} {\hat{a}_{1}}
\newcommand{\adue} {\hat{a}_{2}}
\newcommand{\spiu} {\hat{s}_{0}}
\newcommand{\smeno} {\hat{s}_{1}}
\newcommand{\sv} {\hat{s}_{2}}
\newcommand{\bs} {\hat{a}_1}
\newcommand{\cs} {\hat{a}_3}
\newcommand{\bi} {\hat{a}_{2}}
\newcommand{\ci} {\hat{a}_4}
\newcommand{\sigmap} {\hat{s}_1}
\newcommand{\sigmam} {\hat{s}_2}
\newcommand{\deltap} {\hat{s}_3}
\newcommand{\deltam} {\hat{s}_4}
\newcommand{\gbar} {\bar{g}} 
\newcommand{\rbar} {\bar{r}}
\newcommand{\rS}   {r_{\sigma}}  
\newcommand{\rD} {r_{\delta}}  
\newcommand{\LD}    {\Lambda_\delta}     
\newcommand{\LS}  {\Lambda_\sigma}      
\newcommand{\LSD}    {\Lambda_{\sigma,\delta}}  
\newcommand{\SD}    {\sigma,\delta} 
\renewcommand{\SS}     {\sigma}   
\newcommand{\DD}     {\delta}   
\newcommand{\phimeno}{\phi_{-}}

\newcommand {\chidue } { \chi^{(2)} }
\newcommand{\bsub}{\begin{subequations}}
\newcommand{\esub}{\end{subequations}}
\newcommand{\beq}{\begin{equation}}
\newcommand{\eeq}{\end{equation}}
\newcommand{\beqa}{\begin{eqnarray}}
\newcommand{\eeqa}{\end{eqnarray}}
\newcommand{\beql}{\begin{subequations}\begin{eqnarray}}
\newcommand{\eeql}{\end{eqnarray}\end{subequations}}
\documentclass[pra,aps,amsmath, amssymb,nofootinbib,showpacs,longbibliography]{revtex4-2}
\usepackage{amsmath} 
\usepackage{empheq}
\usepackage{amssymb}
\usepackage{mathbbol}
\usepackage{dsfont}
\usepackage{graphicx}
\usepackage{longtable}
\usepackage{calligra}
\usepackage{array}
\usepackage{multirow} 
\usepackage[table]{xcolor}
\usepackage{float}   
\usepackage{verbatim}  
\usepackage{braket}
\usepackage{color}
\usepackage{framed}
\usepackage{ushort}
\DeclareMathOperator{\tg}{tg}
\begin{document}
\title{Multipartite spatial entanglement generated by concurrent nonlinear  processes}
\author{ Alessandra~Gatti$^{1,2}$}
\affiliation{$^1$ Istituto di Fotonica e Nanotecnologie (IFN-CNR), Piazza Leonardo  Da Vinci 32, Milano, Italy;  
$^2$ Dipartimento di Scienza e Alta Tecnologia dell' Universit\`a dell'Insubria, Via Valleggio 11,  Como, Italy }
\email{Alessandra.Gatti@ifn.cnr.it}
\begin{abstract}
 Continuous variables multipartite entanglement is a key resource for quantum technologies. This works considers the multipartite entanglement generated in separated spatial modes of the same light beam by three different parametric sources: a standard $\chidue$ medium pumped by two pumps,  a single-pump nonlinear photonic crystal, and a doubly pumped nonlinear photonic crystal. These sources have in common the coexistence of several concurrent nonlinear processes in the same medium, which allows the generation of non-standard 3 and 4-mode couplings.  We test the genuine nature of the multipartite entangled states thereby generated in a common framework, using both criteria based on proper bounds for the variances of nonlocal observables and on the positive partial transpose criterion. The relative simplicity of these states allows a (hopefully) useful comparison of the different inseparability tests. 
\end{abstract}
\centerline{Version \today}
\maketitle
\nopagebreak
\section*{Introduction}
In optics,  the most efficient sources of quantum states are nonlinear processes,  such as parametric down-conversion (PDC),  that generate photons in pairs.  
 In the continuous-variable regime this mechanism  naturally leads  to a  bipartite Einstein-Podosky-Rosen  (EPR) entanglement, involving  the phase quadratures of  independent couples of  modes of the radiation field,  or   to  squeezed states, depending  whether   the detection process separates or not the photons in the pair. 
On the other side, photonic states in which entanglement is shared by more than two physical modes are becoming more and more an attractive   resource for continuous variable quantum information. A prominent examples is that of  measurement-based  quantum computation \cite{Raussendorf2001, Menicucci2006}, which requires to generate  multipartite entangled cluster states 
 \cite{Briegel2001, Zhang2006} in a controlled and re-configurable way.  Multipartite entanglement may also  enable  multiparty quantum communication protocols \cite{Armstrong2015}, as  secret quantum sharing  \cite{Hillery1999}, or can be used in quantum metrology for distributed quantum sensing \cite{Shapiro2018}. 
\par 
The traditional method to transform the bipartite entanglement typical of parametric processes into a  multiparty entanglement is  sequential, and is based on  generating several squeezed state by independent PDC processes and then mixing them into a network of passive optical elements  (see e.g. \cite{VanLoock2007, Furusawa2008, Furusawa2013}). Multipartite entanglement may be also  realized by cascaded nonlinearities \cite{Shalm2013}. Alternatively, a powerful approach is the parallel generation of a multipartite entanglement among different light modes that copropagate in the same beam. Continuous variable cluster states have been successfully generated in the spectral structure of the frequency comb of a single optical parametric oscillator \cite{Physer2011,Zhu2021}, or of a synchronously pumped optical parametric oscillator \cite{Roslund2014, Cai2017}.  Spatial encoding , which is naturally attractive,   has been in comparison less explored, with  a recent proposal exploiting an array of  nonlinear waveguides \cite{Barral2020}  interacting with evanescent coupling, which mixes  squeezing and entanglement. 
\par
This work stems from  some recent publications \cite{Gatti2018,Gatti2020b,Gatti2020, Jedr2018,Jedr2020,Brambilla2019}, which showed the possibility of generating  a  multipartite coupling - the prerequisite for multiparty entanglement -  among  separate  {\em spatial}  modes copropagating in the same beam, by properly engineered parametric processes.  The present work  will not deal with the physics of the optical sources,  extensively described in Refs. \cite{Gatti2018,Gatti2020b,Gatti2020, Jedr2018,Jedr2020,Brambilla2019}, but will focus on the  characterization and possibly on the quantification of  the multipartite entanglement thereby produced. 
\begin{figure}[hbt]
\includegraphics[scale=0.65]{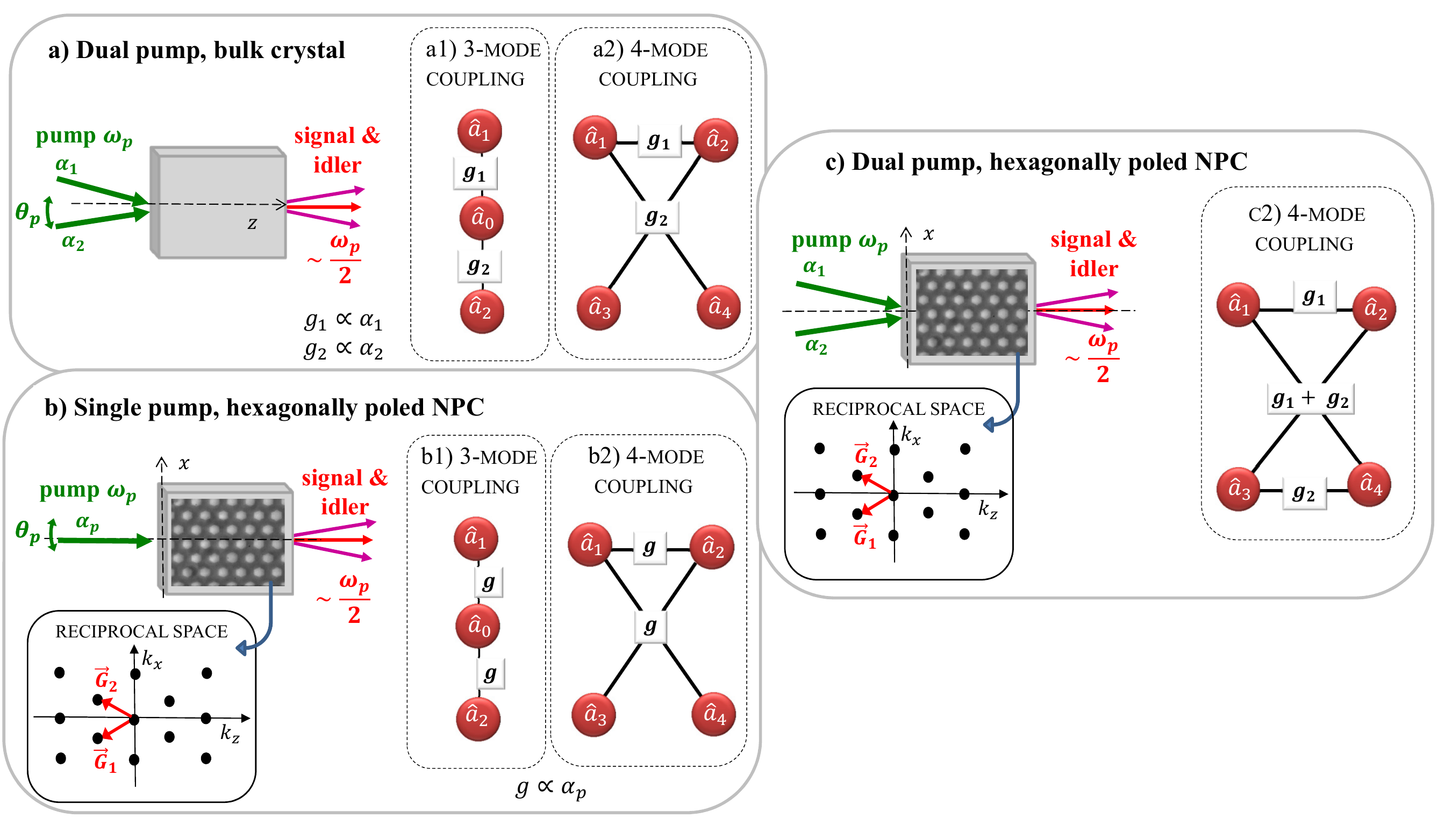}
\caption{Parametric sources of multipartite spatial entanglement. a) Standard  $\chidue$ medium 
pumped by
 two noncollinear beams of amplitudes  $\alpha_1$ and $ \alpha_2$. b)  Hexagonally poled \HEX, pumped by a single pump $\alpha_p$, where two non collinear processes are sustained by  vectors $\vec G_1$ and $\vec G_2$ of the reciprocal lattice. c) Doubly-pumped \HEX, where  four processes coexist.  The graphs describe the multi-mode coupling generated in  specific sets of spatio-temporal modes, in normal conditions [ a1) and b1)]  or under special resonance conditions  [a2), b2) and c2)]  (see text). The couplings $g_j \propto \alpha_j$, while $g \propto \alpha_p$
}
\label{fig_schemes}
\end{figure}
Figure \ref{fig_schemes}  summarizes the optical schemes that will be considered,  along with the states generated by each of them. In   Fig.\ref{fig_schemes}a,  the  medium is a standard $\chidue$ crystal,  but   the beam that feeds the process is engineered  in the form of two slightly non collinear modes, which  determines a transverse modulation of the pumping profile.
In the   case   of Fig.\ref{fig_schemes}b   the pump is a single-mode beam, but the  medium is a nonlinear photonic crystal (NPC) \cite{Berger1998}, whose   $\chidue$   response is artificially modulated according to a two-dimensional poling pattern.
 The common feature of these apparently disparate sources is the coexistence   inside the same medium of two concurrent nonlinear processes,  which 
mutually reinforce during propagation, and the existence of spatio-temporal modes of the fluorescence radiation shared by both processes. 
These  constitute an infinite  set of bright modes \cite{Jedr2018,Jedr2020},  characterized by  the   non-standard 3-mode coupling  schematically depicted by the graphs a1) and b1). 
Under special conditions, a transition to the linear 4-mode coupling of graphs a2) and b2) can take place \cite{Gatti2018,Jedr2018, Gatti2020b,Jedr2020}.  The third example is a mixture of the other two, corresponding to a nonlinear photonic crystal pumped by two noncollinear optical modes. In this case, four nonlinear processes coexist in the same medium, and under particular resonance conditions give rise to the square  coupling in the graph c2) \cite{Gatti2020,Brambilla2019}. 
\par 
The paper has two main parts.
 In the first one, following previous literature \cite{Peres1996, Horodecki1996, Simon2000,Werner2001,Vidal2002, VanLook2003, Adesso2007,Shalm2013,Teh2014, Toscano2015} we set the  machinery for describing continuos variable multipartite entanglement and we establish the  general criteria that will be   used  in the second part  for characterizing our  states. 
 We shall  use two methods: the first one, in the spirit of \cite{Duan2000, VanLook2003}, requires setting proper 
bounds for the  variances of  nonlocal observables  of the system,   that when violated  certify the entanglement of the state. Our inequalities have nothing substantially new compared to others derived and used in existing literature\cite{VanLook2003, Teh2014,Toscano2015}. 
However, they have the advantage of being particularly simple and general, and of having the compact form of Heisenberg-like inequalities [see Eq.\eqref{PPTvar}]. Along with them,  we also suggest a viable strategy to identify the nonlocal observables   best suited to test the inseparability of  a given Gaussian state, based on its  Bloch-Messiah decomposition \cite{Braunstein2005}.   Due to the relative simplicity of our states, we have full access to their covariance matrix, which gives the possibility of using more powerful inseparability tests directly based on the positive partial transpose criterion \cite{Peres1996, Horodecki1996,Simon2000}. This allows a (hopefully) instructive comparison between the two kinds of inseparability tests. 
\section{CV Multipartite entanglement}
\label{sec_general}
Let us start by fixing the formalism and introducing the basic concepts. We consider a system of N optical modes, with bosonic annihilation an creation operators $\hat a_j, \hat a_j^\dagger$  ($j=1,..N$). Hermitian  quadrature operators are defined as
$\hat X_j = \hat a_j + \hat a_j^\dagger$, $\hat Y_j = \frac{1}{i} (\hat a_j - \hat a_j^\dagger)$,  with commutator  $ \left[ \X_j, \Y_k \right] = 2i \delta_{j, k}$. By introducing the  
 2N-dimensional vector
\beq
 \vQ=\left(\X_1, ...,\X_N, \Y_1,  ...,\Y_N \right)^\intercal \, ,
\eeq
the fundamental commutation relations take the compact form:  
\beq
\left[\Q_{\alpha}, \Q_{\beta}\right] = 2i \Omega_{\alpha \beta}
\label{comm}
\eeq
where $\Sy$ is the $2N \times 2N$  {symplectic}  form $\Sy = \begin{pmatrix} \zero_N  & \id_N \\   -\id_N &   \zero_N \end{pmatrix} $, with $ \id_N$  and $\zero_N$ being the  identity and null matrix in N-dimensions \footnote{The specific form of $\Sy$ depends on the order in which operators are arranged into $\vQ$.  This is often defined as $\vQ= (\X_1,\Y_1, \ldots \X_N,\Y_N)$, so that $\Sy$ becomes a  block diagonal matrix, where each block is $\left( \begin{smallmatrix}  0&1\\-1 &0  \end{smallmatrix}\right)$}.  Each state $\hat \rho$ of the system can be  associated with a covariance matrix,  containing  the second order  moments of quadrature operators in symmetric ordering:  
\beq
V_{\alpha \beta} =\llangle \left\{ \dQ_\alpha, \dQ_{\beta}  \right\} \rrangle =\Tr\left \{ \half (\dQ_\alpha \dQ_{\beta}  +  \dQ_{\beta} \dQ_\alpha)  \hat \rho  \right\} 
\eeq
where $\dQ_j = \Q_j -\langle \Q_j \rangle$. Gaussian states, in particular, are uniquely determined by their covariance matrix, a part from a displacement in phase space, inessential for the entanglement. A legitimate covariance matrix must be real, symmetric and must satisfy the condition: 
\beq
\V + i \Sy \ge 0
\label{positivity}
\eeq
meaning that  the Hermitian matrix $ \V + i \Sy $ has non-negative eigenvalues (the same clearly holds true for the complex conjugate $\V - i \Sy $, and for the covariance $\V$ itself). This condition is a straightforward consequence of the positivity of the density operator  $\hat \rho$,  and of the commutation relation \eqref{comm},  because it ensures that any operator of the form    $\hat \xi= \sum_\alpha c_\alpha \dQ_\alpha $, where $c_\alpha$ are complex coefficients, satisfies  
$$
\langle \hat  \xi^\dagger \hat \xi \rangle = 
\sum_{\alpha, \beta} c_\beta^*  c_\alpha \langle  \dQ_\alpha   \dQ_{\beta} \rangle =
\sum_{\alpha, \beta} c_\beta^*  c_\alpha \left\langle  \left\{ \dQ_\alpha, \dQ_{\beta} \right\}   +\half      \left[ \Q_\alpha, \Q_{\beta} \right]   \right\rangle 
= \vect{c}^\dagger \left(\V + i \Sy \right) \vect{c} \ge 0 
$$
for any 2N-dimensional vector of complex numbers 
 $\vect{c}^\dagger = (c_1^*,..,c^*_{2N})$, including  the eigenvectors of  the matrix $\V + i \Sy $, whose eigenvalues are therefore non-negative.  The condition \eqref{positivity}   is often expressed  in the equivalent form  (see e.g.  \cite{Adesso2007}). 
\beq
 \mathrm{Eigen}_+ \left[   i \Sy \V \right]   \ge 1
\label{positivity2}
\eeq
where $ \mathrm{Eigen}_+ [ \cdot]    $ denotes the positive eigenvalues  of $ i \Sy \V $, which form the so-called  {\em symplectic spectrum}   $\vect{\nu} = \{\nu_1,\nu_2 \ldots \nu_N \} $ of  the covariance matrix. The inequality \eqref{positivity} implies that for a  legitimate covariance matrix  $\nu_i \ge 1$ , $ \forall i=1,\ldots,N$ 
 (see appendix \ref{app_Williamson} for details).  
\par
The inequalities \eqref{positivity} and \eqref{positivity2}
can be considered as a general expressions of the Heinsenberg uncertainty relations,  which bound the variances of  pairs of observables.  If we focus on the simplest type of  non-local observables, i.e.   linear combinations of  the quadrature operators of the modes
\beq
\begin{aligned}
&\hat \eta (\vd) = \sum_\alpha d_\alpha \Q_\alpha, \\ &    \hat \eta (\vd') = \sum_\alpha d'_\alpha \Q_\alpha, 
\end{aligned}
\label{eta}
\eeq
 where $\vd= (d_1,d_2 \ldots d_{2N})^\intercal$, $ \vd'= (d'_1,d'_2 \ldots d'_{2N})^\intercal$ are vectors of {\em real} coefficients,    their variances must satisfy the following Heisenberg bound: 
\beq
\begin{aligned}
\llangle \delta \hat \eta^2  (\vd) \rrangle  + \llangle \delta \hat \eta^2  (\vd^\prime) \rrangle &  \ge 2 
\sqrt{ \llangle \delta \hat \eta^2  (\vd) \rrangle} \sqrt{    \llangle \delta \hat \eta^2  (\vd^\prime) \rrangle}
\\
& \ge \left| \llangle  \left[\delta \hat \eta  (\vd),  \delta \hat \eta  (\vd^\prime)    \right] \rrangle \right|
\label{Heisenberg}
\end{aligned}
\eeq
As well known, these inequalities are  a purely mathematical consequence of the positivity of the density operator $\hat \rho$, and of the fact that $\hat \eta (\vd) $ 
and $\hat \eta (\vd^\prime) $ are Hermitian operator (observables).  Indeed it is based on the Cauchy–Schwarz inequality:   
$ {\llangle \delta \hat \eta^2  (\vd) \rrangle} {   \llangle \delta \hat \eta^2  (\vd^\prime) \rrangle} \ge 
\left| \llangle \delta \hat \eta  (\vd)  \delta \hat \eta  (\vd^\prime)   \rrangle \right|^2   $, and on the inequality 
$\left| \llangle \delta \hat \eta  (\vd)  \delta \hat \eta  (\vd^\prime)   \rrangle \right|  =    \left|   \vd^\intercal ( \V  + i \Sy )\vd^\prime  \right|
\ge  \left|  \vd^\intercal \Sy\,  \vd^\prime \right| = \half   \left| \llangle  \left[\delta \hat \eta  (\vd),  \delta \hat \eta  (\vd^\prime)    \right] \rrangle \right|$, which in turn is a straightforward consequence of the fact that $\vd$ 
and $\vd'$ are real vectors. \\
 Notice also that the bound for the product of variances  is  stronger than the one for the sum , because  the inequality in the first line of the formula\eqref{Heisenberg} holds  strictly   unless the two variances  are equal. 

\subsection{The PPT criterion in phase space}
We consider now the problem of separability of the state with respect to a given bipartition. Let us consider a partition of the N modes into two subgroups 
$A = \{j_1,j_2 \ldots j_k\}$ (Alice) and $ B= \{m_1, m_2 \ldots m_{N-k} \}=A^C$ (Bob).  We  call {\em A-separable}  those states  that  can be written in the form  $\hat \rho = \sum_n  P_n  \hat \rho_n^{(\mathrm{A})}   \otimes \hat \rho_n^{(\mathrm{B})}$, where  $P_n \ge 0$, and $\hat \rho_n^{(\mathrm{A})}$ and $\hat \rho_n^{(\mathrm{B})}$ are density operators on Alice and Bob subspaces. \\
The  Peres-Horodecki \cite{Peres1996, Horodecki1996} Positive Partial Transpose (PPT) criterion establishes  that for  any A-separable state  the operator  
 $\hat \rho^{ \mathsf{\, \scriptscriptstyle PT(A)}} $
obtained by partial transposition of  its density operator with respect to the degrees of freedom of subsystem A is still a legitimate density operator. The existence of  a negative partial transpose density operator  is thus a sufficient criterion for assessing the A-entanglement of the state, and becomes also necessary for   the $\{j_1\} \times \{m_1,..m_{N-1}\}$ partitions of Gaussian states \cite{Simon2000,Werner2001}. For the continuous variable systems of interest for this work, an elegant and useful translation of the PPT criterion to phase-space has been performed by Simon \cite{Simon2000}, who showed that in phase-space  the partial transposition corresponds to a mirror reflection of all the Y-quadratures of  Alice's modes.  
Mathematically,  this amounts  to a  transformation  of the covariance matrix of the state, by means of the unitary matrix   
\beq
\PT_A= \mathrm{diag}\{1,1, \ldots \underset{ \underset{N+j_1}{\downarrow}} {-1}\ldots \underset{ \underset{N+j_2}{\downarrow}} {-1} \ldots \underset{ \underset{N+j_k}{\downarrow}} {-1}\}  
\label{Mirror}
\eeq
 which changes the sign of the Y-quadratures of modes $j_1,j_2\ldots j_k$ of the set A. 
The PPT criterion  for continuous variables  \cite{Simon2000,Werner2001}  establishes that  for any  A-separable state the matrix 
\beq
\V^{ \mathsf{\, \scriptscriptstyle PT(A)}} = \PT_A \V \PT_A
\label{VPT}
\eeq
is still a legitimate covariance matrix, which implies
\beq
 \V^{ \mathsf{\, \scriptscriptstyle PT(A)}}  + i \Sy \ge 0 
\label{VPTpos}
\eeq
or,  alternatively 
\beq
\mathrm{Eigen}_+ \left[   i \Sy \V^{ \mathsf{\, \scriptscriptstyle PT(A)}}  \right] \ge  1
\label{VPTsympl}
\eeq
The conditions \eqref{VPTpos} or   \eqref{VPTsympl}  provide  powerful  means to  test the entanglement of the state with respect to the partition $A$:   whenever a negative eigenvalue of $\V^{ \mathsf{\, \scriptscriptstyle PT(A)}} +i \Sy$  appears (or a symplectic eigenvalue of $\V^{ \mathsf{\, \scriptscriptstyle PT(A)}}$  is  smaller than 1),  then $\hat \rho^{ \mathsf{\, \scriptscriptstyle PT(A)}} $ is not a physical state, which implies that  $\hat \rho$ is not A-separable. Clearly, such a   test requires the access to the full covariance matrix  of the state, which is often not a viable route in experiments because of the  large number of measurements required. 
 \subsection{Criteria based on variances of nonlocal observables} 
\label{sec_PPTvar}
Simon \cite{Simon2000} proposed  a  somehow more  accessible use of the  PPT criterion, that was subsequently applied to various examples of multipartite   CV entanglement   \cite{VanLook2003, Shalm2013,Teh2014}, and systematically generalized  by the work in \cite{Toscano2015}.  We propose here a simpler  version of the general criteria described in \cite{Toscano2015}, sufficient for characterizing the states of interest in this work. 

The Heisenberg inequalities  \eqref{Heisenberg} hold for any physical state. A-separable states conversely  impose stronger bounds, originating from their  positive partial transpose.   Because of the equivalence between the partial transposition of the density operator and the mirror reflection of  the Y-quadratures of Alice,   the variances of each pair  of linear combinations of mode quadratures $\hat \eta (\vd)$ and $\hat \eta (\vd^\prime)$  satisfy the relation: 
\begin{align}
\llangle \delta \hat \eta^2  (\vd) \rrangle_{\hat \rho}  + \llangle \delta \hat \eta^2  (\vd^\prime) \rrangle_{\hat \rho} =
 \llangle \delta \hat \eta^2  (\PT_A \vd) \rrangle_{\hat \rho^{ \mathsf{\, \scriptscriptstyle PT(A)}} }  + \llangle \delta \hat \eta^2  (\PT_A \vd^\prime) \rrangle_{\hat \rho^{ \mathsf{\, \scriptscriptstyle PT(A)}} }
 \label{equiv}
\end{align}
where $ \llangle \cdot \rrangle_{\hat w} = \Tr\{ \cdot  \hat w  \}$,  $\hat \rho^{ \mathsf{\, \scriptscriptstyle PT(A)}}$  is the partial transpose (with respect to Alice's set) of the density operator, 
and $\PT_A$ is the mirror reflection described by Eq.\eqref{Mirror}.  For A-separable states $\hat \rho^{ \mathsf{\, \scriptscriptstyle PT(A)}}$ is a valid density operator, so that the r.h.s of Eq.\eqref{equiv} must satisfy a Heisenberg bound analogue to that of Eq.\eqref{Heisenberg}, which is a purely algebraic consequence of the positivity of the density operator. Therefore, in A-separable states the variances of observables are subject to the additional bound 
\bsub
\begin{align}
\llangle \delta \hat \eta^2  (\vd) \rrangle  + \llangle \delta \hat \eta^2  (\vd^\prime) \rrangle 
&\ge 2  \left|  \vd^\intercal  \PT_A \Sy\, \PT_A \vd^\prime \right| 
 \label{PPTsum1}
\end{align}
In a similar way, the product of variances in A-separable states are constrained to 
\begin{align}
\sqrt{ \llangle \delta \hat \eta^2  (\vd) \rrangle  \llangle \delta \hat \eta^2  (\vd^\prime) \rrangle }
&\ge   \left|  \vd^\intercal  \PT_A \Sy\, \PT_A \vd^\prime \right| 
 \label{PPTprod1}
\end{align}
\label{PPT1}
\esub
Actually, equations \eqref{PPTsum1} and \eqref{PPTprod1} are just an example of the general criteria described in \cite{Toscano2015}, which involve any kind of functional of nonlocal observables.
 From the definition of $\hat \eta$  in Eq.\eqref{eta}, we notice that 
$  \left[\delta \hat \eta  (\PT_A \vd),  \delta \hat \eta  (\PT_A\vd^\prime)    \right] = 2 i    (\PT_A\vd )^\intercal  \Sy\, \PT_A \vd^\prime   
$, so that  Eqs.\eqref{PPTsum1} and \eqref{PPTprod1}  can be recast in the more expressive form:
\beq
\begin{aligned}
\llangle \delta \hat \eta^2  (\vd) \rrangle  + \llangle \delta \hat \eta^2  (\vd^\prime) \rrangle 
& \ge  2 \sqrt{ \llangle \delta \hat \eta^2  (\vd) \rrangle  \llangle \delta \hat \eta^2  (\vd^\prime) \rrangle } \\
& \ge \left| \llangle  \left[\delta \hat \eta  (\PT_A \vd),  \delta \hat \eta  (\PT_A\vd^\prime)    \right] \rrangle \right| 
\label{PPTvar}
\end{aligned}
\eeq
 This inequality has the same form as the Heisenberg relation \eqref{Heisenberg},  but  the  lower bound is determined  by the commutator  between the mirrored variables.   Thus we can say that  the uncertainties of  observables in separable states are constrained  not only by  their commutator, as in   standard Heisenberg uncertainty relations, but also by the commutators of the mirrored observables. The inequality \eqref{PPTvar} has the further  advantage  of not depending on the  
numerical value  of the commutator  (often a source of confusion because of the different definitions of $\hat X, \hat Y$  adopted by various authors), 
and will be  systematically used in the next sections of this work.   
\par
If the nonlocal  variables  $\hat \eta ( \vd )$ and $\hat \eta (\vd^\prime)$ are properly chosen, Eq.\eqref{PPTvar} jointly with the Heisenberg inequality \eqref{Heisenberg}, may provide  a stronger bound than Eq. \eqref{Heisenberg} alone,  obeyed by all the states.
A simple example is that of Duan criterion \cite{Duan2000} for bipartite entanglement between two bosonic modes $\hat a_1$ and $\hat a_2$. Let us consider the nonlocal variables $  \hat \eta= 
\frac{1}{\sqrt{2}} \left( s \hat X_1 - \frac{1}{s} \hat X_2 \right)$  and   $  \hat \eta'= \frac{1}{\sqrt{2}} \left( s \hat Y_1 + \frac{1}{s} \hat Y_2 \right)$, where $s$ is a real number. In  any state,  the sum of their uncertainties cannot be below the Heisenberg bound  $ \left| \left[  \hat \eta, \hat \eta' \right]    \right|  =  |s^2 -  \frac{1}{s^2} |$, which vanishes for s=1. Conversely, separable states needs to respect the stronger bound represented by the commutator of mirrored variables
$ \llangle \delta \hat \eta^2  \rrangle  + \llangle \delta \hat \eta'^2 \rrangle  \ge \half \left| \left[   s\hat X_1 - \frac{1}{s} \hat X_2,  s \hat Y_1 - \frac{1}{s} \hat Y_2 \right]    \right| = s^2 +  \frac{1}{s^2} $.  Therefore, violation of this bound is a sufficient condition for the entanglement of any bipartite continuous variable state (which becomes also  necessary  for Gaussian states \cite{Duan2000}). 
\par
In the general case of a N-mode state,  
negation of a  bound  of the form described by  Eq.\eqref{PPTvar} jointly with Eq.\eqref{Heisenberg}],  is sufficient to   rule out the A-separability of the state.  The demonstration of   a {\em genuine} N-party entanglement  requires then verifying  the inseparability of the state  with respect to each of the  possible partitions of the N modes, whose number is $2^{N-1}-1$. 
\footnote{A simple demonstration that the number of bipartitions of N modes is  $2^{N-1} -1$ can be done by induction. If the number of partitions of N-1 modes is F(N-1), then the number of partitions of N modes is $F(N)= 2F(N-1) +1$, because by adding a new mode, say $j_{N}$, each grouping  $A\times B$ of the ${j_1,...j_{N-1}}$ modes generates the 2 new partitions $\{A,j_{N}\} \times B$ and ${A} \times \{B,j_{N}\}$. Moreover there is the additional partition ${j_{N}}\times \{j_1,..j_{N-1}\}$. Clearly $F(2)=1$, and by using the above recurrence relation $F(N) = \sum_{k=0}^{N-2} 2^k = 2^{N-1} -1$.}
To be truly precise, as observed by \cite{Shalm2013, Teh2014}, such a test is  able to prove the full N-party inseparability of the state and  not its genuine  N-party  entanglement, because the state could be a mixture of A-separable, A'-separable ...etc. states.  However, for  pure states the two concepts luckily coincide, and thus in the following we shall simply identify genuine N-party entanglement of a state with its inseparability with respect to any bipartition. 
\subsection{Gaussian states and Bloch Messiah reduction}
\label{sec_BlochMessiah}
The inequalities  \eqref{PPTvar}   provide a viable strategy to test the separability of the state with respect to any bipartition, especially for pure Gaussian states for which the Bloch Messiah reduction  can be performed.  According to the Bloch-Messiah theorem \cite{Braunstein2005},  any N-mode pure  Gaussian state can be decomposed into N independent squeezed states followed by linear passive transformations, i.e. transformations that do not mix creation and destruction operators (namely beam splitters and phase rotations).  In terms of bosonic operators, it is always possible to find a  $N$x$N$ unitary matrix  $\UB$, such that $\UB\UB^\dagger =  \UB^\dagger \UB   = \id_N$,  and 
\beq
\begin{pmatrix} 
\hat a_1 \\

\vdots \\
\hat a_N 
\end{pmatrix}  = \UB  \begin{pmatrix} 
\hat s_1 \\

\vdots \\
\hat s_N 
\end{pmatrix}
\eeq
where $\hat s_j$ are   bosonic operators in    independent squeezed states (eventually in the vacuum state). The  same decomposition transforms the quadrature operators as 
$ \vQ = \SB  \vQ_s$, where 
$\SB = \begin{pmatrix} \Real (\UB)   & - \Imag (\UB)      \\[0.2em] \Imag (\UB)     &   \Real (\UB) 
  \end{pmatrix} $ is a symplectic matrix, and  the vector $ \vQ_{s} = ( \hat X_{s_1}, \ldots ,  \hat Y_{s_N} )^\intercal $ contains the quadrature operators of the squeezed modes, which are  characterized by  conjugate  variances 
 $ \langle \delta \hat X_{s_j}^2 \rangle= \sigma_{j}$  and $ \langle \delta \hat Y_{s_j}^2 \rangle=1/ \sigma_{j}$. 
By inverting    $\SB $,  the   quadrature of the squeezed modes  can be expressed  as linear combinations  of those of the  original modes: then,  the  squeezed quadratures (say $\hat Y_{sj}$ , $j=1..k$, with $k \le N$) straightforwardly provide  a set of  nonlocal operators whose variances are below shot-noise, and potentially vanish in the limit of large squeezing parameters. Notice that the nonlocal variables obtained in this way commute pairwise by construction, because the squeezed modes are independent.  Therefore, the only bounds which need to be negated in order to demonstrate entanglement  are those defined by the Heisenberg-like  inequalities for mirrored observables  in Eq.\eqref{PPTvar}.  We shall see applications of this procedure in the next sections. 
\par

Clearly, knowledge of the Bloch-Messiah decomposition  permits also to calculate the covariance matrix, which for a pure state can can be  written in the form
\beq
\begin{aligned}
\V & = \SB  \V_\mathrm{sq}  \SB^\intercal \qquad &
\V_\mathrm{sq} =\left( \begin{smallmatrix}  
 \sigma_1 & 0& & & \ldots   & &  &0 \\
0  &  \sigma_2 & 0& & & \ldots    &  &0 \\
&  & \ddots      &    &  & & & \vdots \\
 & &  & \sigma_N  &  & &  & \\
 & &      &                & {\sigma_1}^{-1} & & &\\
\vdots&  &         &              &&  \ddots    & \\
0 & & \ldots   & &   & & & {\sigma_N^{-1}}
                \end{smallmatrix} \right)
\end{aligned}
\label{Vcalc}
\eeq


\section{Doubly pumped bulk crystal, or single-pump \HEX : 3-mode coupling} 
\label{sec_Tripartite}
Let's start with the case of two concurrent nonlinear processes  in the standard configuration, i.e. away from the resonances that will be studied in the next section. 
\par 
Specifically, we consider the scheme   of Fig.\ref{fig_schemes}a,    in which  two pump beams propagate non-collinearly in a bulk $\chidue$ crystal  and originate two intersecting  families of   fluorescence cones (see \cite{Gatti2020b, Jedr2020}  for details).  A second example  is that of   a  2D nonlinear photonic crystal   pumped by a single beam in  Fig.\ref{fig_schemes}b:  in this case, 
 two nonlinear  processes 
coexist in the same medium  because phase-matching can be simultaneously satisfied by means of 
two  non-collinear   vectors ($\vec G_1$ and $\vec G_2$ in the figure) of the  reciprocal lattice of the  2D poling pattern \cite{Jedr2018,Gatti2018}.  In both examples, 
the possibility of having a multiparty entanglement arises from the existence of spatio-temporal  modes of the down-converted light { shared}  by both processes, which host photons generated indistinguishably by process 1 or by process 2. When a photon is created in such a mode, its twin  appears  in  either  of two paired modes. This mechanism naturally leads to a tripartite entangled state. In practice, the triplets of entangled modes constitute an infinite set of bright modes, which appear as hot-spots on a background of ordinary (i.e. 2-mode) parametric fluorescence: their position and frequency are determined by the requirement that phase matching of both processes, i.e. the conservation of the momentum in the microscopic process of  pair creation,  is simultaneously satisfied for  the same space-time mode of the fluorescence radiation \cite{Gatti2018, Jedr2018, Jedr2020,Gatti2020b}.
\par 
Let us focus on a specific triplet of modes. Let    $\azero$ be   the photon destruction operator of the shared mode,  and $\auno$ and $\adue$  those  of the modes coupled to it via process 1  and  2, respectively. Their evolution equations along the sample can be found in \cite{Gatti2020b} in the case of a doubly pumped bulk crystal, or in \cite{Gatti2018} for the single-pump NPC. For perfect phase-matching,   such equations  can be  written 
as $ \hat a_j (z)=  e^{ -\frac{i} {\hbar} \hat {\cal P} z}\hat a_j (0)  e^{ \frac{i} {\hbar} \hat {\cal P} z}$,  where   $  e^{ \frac{i} {\hbar} \hat {\cal P} z}$ is the   unitary  z-evolution operator, and 
the  
 longitudinal momentum operator  is 
\beq
\hat {\cal P} =- i \hbar \left( g_1 \azero^\dagger a_1^\dagger + g_2 \azero^\dagger \adue^\dagger - g_1^* \azero \auno -g_2^* \azero \adue \right) .
\label{momentum}
\eeq
Here $g_1$ and $g_2$ are the parametric coupling strengths  of each process. In  the dual pump case they are proportional to the  complex amplitudes  $\alpha_1$ and $\alpha_2$ of the two pump waves, while for the  single-pump NPC   $g_1=g_2:= g \propto \alpha_p$ . Alternatively, in the picture where the states evolves  along the medium 
\beq 
| \psi (z) \rangle =  e^{\frac{i}{\hbar} \hat {\cal P} z} | \psi_{\mathrm in} \rangle 
= e^ { \left( g_1 \azero^\dagger a_1^\dagger + g_2 \azero^\dagger \adue^\dagger - h.c. \right)  }  |\psi_{\mathrm in} \rangle 
\label{3state}
\eeq

\begin{figure}[ht]
\includegraphics[scale=0.6]{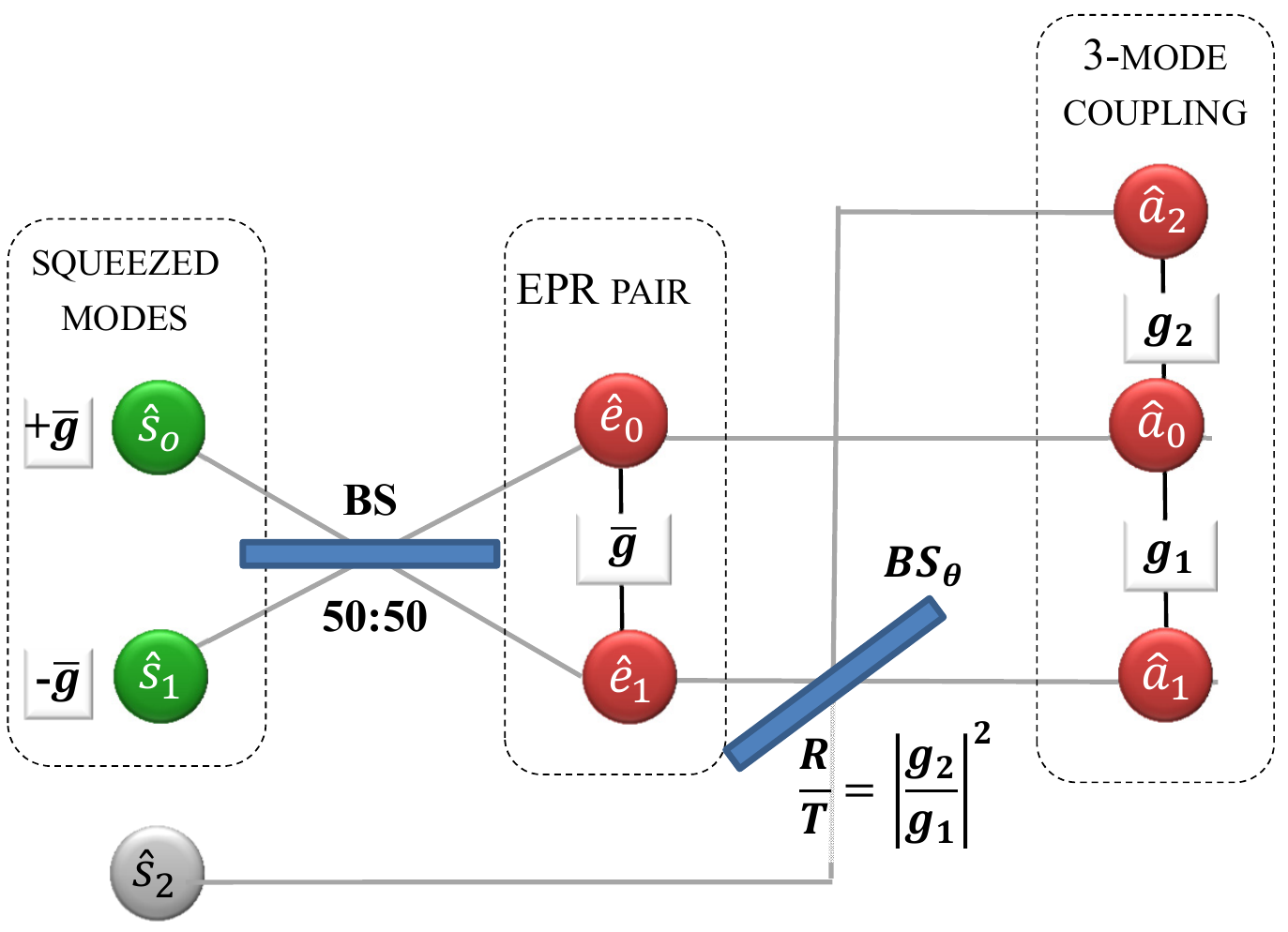}
\caption{Bloch-Messiah decomposition  of the tripartite entangled state generated by two concurrent parametric processes of strengths $g_1$ and $g_2$: $\spiu$ and $\smeno$ are  squeezed  modes, with opposite squeezing parameters $\pm \gbar z$, where $\gbar = \sqrt { |g_1|^2 + |g_2|^2}$. $BS$ is a balanced beam-splitter, while $BS_\theta$  has a splitting ratio  $R/T=  {|g_2|^2}/{ |g_1|^2} $.
}
\label{fig_equiv3}
\end{figure}

In \cite{Gatti2020b} it was shown that the dynamics could be decomposed into a single parametric process  generating a pair of EPR entangled modes, followed by a beam splitter 
that mixes  one of the EPR modes with an independent mode in an arbitrary input state.   In this work we prefer the standard Bloch-Messiah decomposition in terms of squeezed modes. To this end,   the phase rotations $\auno \to \auno e^{i \arg [{g_1}]}$ and  $\adue \to \adue e^{i \arg [{g_2}]}$ are first performed: since these are local operations that do not affect the entanglement of the state, they will be neglected in what follows. Next, 
we consider the   transformation 
\beq
\begin{pmatrix}
\azero \\
\auno \\
\adue 
\end{pmatrix}   = 
\frac{1}{\sqrt 2}
\begin{pmatrix} 1  &  1 & 0 \\[0.1em]
 \cos \theta &  {-\cos \theta}   & -{\sqrt 2}  \sin \theta \\[0.1em]
{\sin \theta} &{-\sin \theta } &  {\sqrt 2} \cos \theta 
\end{pmatrix} 
\begin{pmatrix}
\spiu \\
\smeno \\
\sv
\end{pmatrix} = \UB
\begin{pmatrix}
\spiu \\
\smeno \\
\sv
\end{pmatrix}
\label{uzero}
\eeq
where $ \spiu, \smeno, \sv$ are independent bosonic operators and 
\beq 
\tg \theta =\frac{ |g_2|}{ |g_1|  }  .
\label{theta}
\eeq   It can be easily recognized that it corresponds to the action of a 50:50 beam splitter on modes 0 and 1, followed by a beam-splitter with $R= \sin^2 \theta$,  $T=\cos^2\theta$ acting  on modes 1 and 2. By calling $ \hat { \cal U}_B $    the generator of such  transformation,  and  applying it to  to the momentum 
operator in  Eq.\eqref{momentum}, one easily finds that 
\begin{align}
\hat { \cal U}_B^\dagger  e^{ \frac{i} {\hbar} \hat {\cal P} z}  \hat { \cal U}_B= e^{ \frac{\gbar}{2}  \left(  \spiu^\dagger \spiu^\dagger  -  h.c. \right)z} 
 e^{- \frac{\gbar}{2}   \left ( \smeno^\dagger \smeno^\dagger  - h.c. \right)z} 
=  \hat { \cal S}_0  (\gbar z)  \otimes  \hat { \cal S}_1  (-\gbar z) 
\label{Tri-red}
\end{align} 
where $\hat { {\cal S}}_j  (r) = \exp{ (\frac{r}{2}  \hat s_j^\dagger \hat s_j^\dagger -h.c.)}$ is the single-mode  {\em squeeze} operator for mode $j$,  and 
\beq
\gbar = \sqrt{|g_1|^2 + |g_2|^2}. 
\label{gbar}
\eeq 
In this way, 
 the z-evolution  has been reduced to   the  action of two independent squeezers with  opposite squeezing parameters $\pm \bar r = \pm \gbar z$,  acting on modes $0$ and $1$ (mode 2 is not squeezed), followed by the passive transformation  described by  Eq.\eqref{uzero}: 
$e^{ \frac{i} {\hbar} \hat {\cal P} z} = \hat { \cal U}_B \hat { \Sigma}_0  (\gbar z) \hat { \Sigma}_1 (-\gbar z)  \hat { \cal U}_B^\dagger $.  When applied to  a vacuum input state, the rightmost operator $ \hat  { \cal U}_B^\dagger$ has no effect, and the overall dynamics can be decomposed according to the schematic in Fig.\ref{fig_equiv3}. 
We observe that for the doubly-pumped  crystal the effective squeezing parameter $\rbar= \gbar z \propto \sqrt{ |\alpha_1|^2 + \alpha_2|^2} $  is  proportional to the square root of the total energy. Thus the level of squeezing  coincides with  what would be obtained  by injecting   in the same material a single pump with the same total energy. 
 Conversely,  for the single-pump \HEX ,   $\rbar= \sqrt 2 \, g \propto \sqrt 2 |\alpha_p|$, i.e  there is a net $\sqrt  2$ increase of the amount of squeezing/gain in the 2D poled material, due to the coherent superposition of the two concurrent  nonlinear processes. 
\par
Clearly, Fig. \ref{fig_equiv3} can be read also in the inverse way: when a a member of an EPR pair with squeezing parameter $r=\gbar z$ is mixed with a  vacuum input on a beam-splitter of transmission $T=\cos^2 \theta$ , the result is the tripartite entangled state in the figure, obtained from the action of the momentum \eqref{momentum}, with $g_1= \gbar \cos \theta$ and $g_2= \gbar \sin \theta$. 
\subsection{Tripartite entanglement}
We are now in conditions of applying the entanglement criteria outlined in  sections \ref{sec_PPTvar} e \ref{sec_BlochMessiah}. 
Provided that  squeezing is  along the Y-quadrature for  $\spiu$  while  it is along the X-quadrature  for $\smeno$, inversion of the matrix \eqref{uzero} permits to identify the nonlocal variables with sub-shot noise fluctuations. These are 
\begin{align}
\hat \eta ( \vd) & 
= \frac{1}{\sqrt 2} \left(  \X_0 - \hat X_1  \cos \theta -  \X_2 \sin \theta \right)  = e^{-\gbar z}  \hat X_{s_1}^\mathrm{in}
\label{etah} \\
\hat \eta ( \vd')  & 
= \frac{1}{\sqrt 2} \left(  \Y_0 +  \hat Y_1  \cos \theta+  \Y_2 \sin \theta \right)  = e^{-\gbar z}  \hat Y_{s_0}^\mathrm{in} 
\label{etag} 
\end{align}
where,  according to the formalism developed in Sec.\ref{sec_general},  $\vd =  \frac{1} {\sqrt{2}}  (1, -\cos \theta, -\sin \theta,0,0,0)^\intercal$, and $\vd' =  \frac{1} {\sqrt{2}}  (0,0,0, 1,\cos \theta, \sin \theta)^\intercal$. 
 The  two variables commute $[\hat \eta ( \vd),\hat \eta ( \vd')]=0$, so that there is no lower Heisenberg bound \eqref{Heisenberg} to their  variances. Indeed for vacuum input
\beq 
\langle \delta \hat \eta ^2 ( \vd)\rangle = \langle \delta \hat  \eta^2 ( \vd') \rangle = e^{-2 \gbar z}  \to 0 
\label{etavar}
\eeq 
 for $\gbar z \gg1 $. Let us now check the bounds that must be obeyed by separable states. The bi-partitions of three modes  are those corresponding to the 3 possible choices  of a single mode with respect to the other two. Let us check them one by one, by applying the separability criterion of Eq.\eqref{PPTvar}. \\

\paragraph*{Separability of the shared mode $\azero$.} 
The operation of partial transposition with respect to mode 0  corresponds in phase-space to  the application of the mirroring matrix 
$ \PT_{0} = \mathrm{diag} \{ 1, 1,1, -1,1,1\} $ that inverts the sign  of  $ \Y_0$. Clearly, this has effect only  on $ \vd' $ $\to \Gamma_0 \vd'= \frac{1} {\sqrt{2}}  (0,0,0, -1,\cos \theta, \sin \theta)^\intercal$. Any state separable with respect to mode $0$  must respect the following bound
\beq
 \langle \delta\hat \eta^2 ( \vd) \rangle + \langle \delta \hat \eta ^2 ( \vd')\rangle   \ge \left|  [\hat \eta ( \PT_0 \vd'),\hat \eta ( \PT_0  \vd)] \right| =2 
\label{b0t}
\eeq
which is violated  by our  state for any finite amount of squeezing $\gbar z >0$. Therefore we conclude that mode 0 is never separable from the other two, as should be rather intuitive from the graph of the state in Fig.\ref{fig_equiv3}. \\

\begin{figure}[ht] 
\includegraphics[scale=0.6]{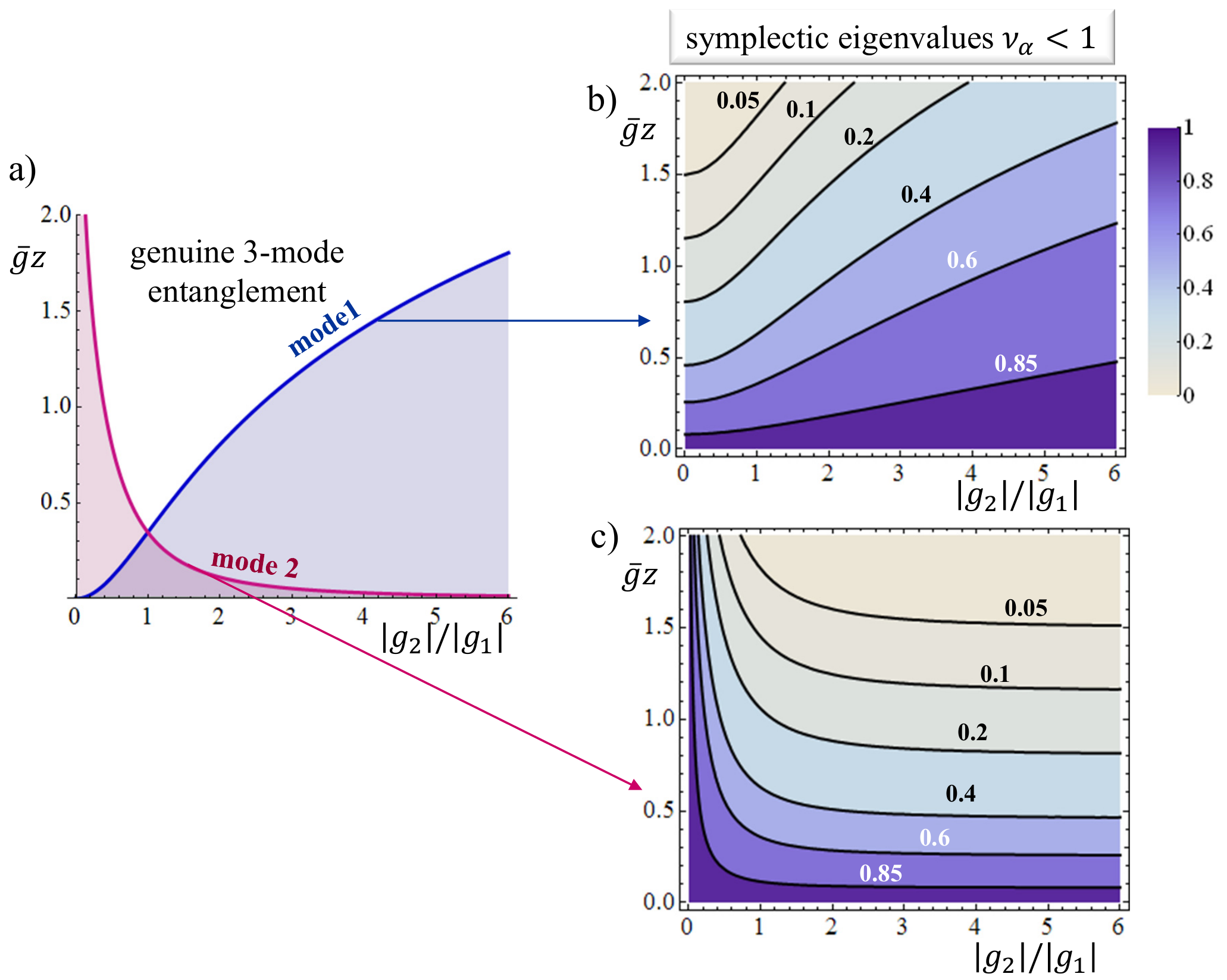}
\caption{Entanglement of the 3-mode state \eqref{3state}. a) Sufficient criteria   based on the violation of  Heisenberg-like inequalities: mode 0 is never separable, while  the entanglement of mode 1 and 2 is verified in the  region above  the blue and the red curve, respectively,  according to   \eqref{var1}  and \eqref{var2}.   b)  and c): Contour plots of the  smallest symplectic eigenvalue $ \nu_{-}^{(1)}  $   and  $ \nu_{-}^{(2)}  $  of the partial transpose with respect to mode 1 and 2, respectively, showing that actually entanglement is always present, unless trivially  $g_1$ or $g_2$ vanish.  }  
\label{fig_3ent} 
\end{figure}

\paragraph*{Separability of mode $\auno$.} 
Partial transposition with respect to mode 1  now  transforms 
$ \vd'  \to \Gamma_1 \vd' $ $= \frac{1} {\sqrt{2}}  (0,0,0, 1, -\cos \theta, \sin \theta)^\intercal$. The bound that must  be respected by 1-separable states is 
\beq
\begin{aligned}
 \langle \delta \hat \eta^2 ( \vd) \rangle + \langle \delta \hat \eta ^2 ( \vd')\rangle  &  \ge  \left| \left [\hat \eta ( \PT_1 \vd'),\hat \eta ( \PT_1  \vd)\right] \right| \\
&= 
 \half \left|  \left [\X_0 - \hat X_1  \cos \theta -  \X_2 \sin \theta   , \Y_0 -  \hat Y_1  \cos \theta+  \Y_2 \sin \theta   \right] \right| \\
& =2 \cos^2 \theta 
\label{b1t}
\end{aligned} 
\eeq
When applied to our  tripartite state, using Eq.\eqref{etavar}, a sufficient condition for inseparability of mode 1 is 
\beq
\gbar z >   \ln{ \sqrt{1 +    \left| g_2/g_1\right|^2}}  
\label{var1}
\eeq
which  is shown in Fig.\ref{fig_3ent}a by the blue curve.  
The condition  becomes more and more demanding as   $g_1 \to 0$:  when mode 1 is weakly coupled, more squeezing is required for verifying its entanglement.  \\

\paragraph*{Separability of mode $\adue$.} 
In this case
$ \vd'  $ $\to \Gamma_2 \vd' = \frac{1} {\sqrt{2}}  (0,0,0, 1, \cos \theta, -\sin \theta)^\intercal$. States separable with respect to mode 2 must satisfy the inequality 
\beq
\begin{aligned}
 \langle \delta \hat \eta^2 ( \vd) \rangle + \langle \delta \hat \eta ^2 ( \vd' )\rangle  &  \ge  \left| \left [\hat \eta ( \PT_2 \vd'),\hat \eta ( \PT_2  \vd)\right] \right| \\
&= 
 \half \left|  \left [\X_0 - \hat X_1  \cos \theta -  \X_2 \sin \theta   , \Y_0  +\hat Y_1  \cos \theta -  \Y_2 \sin \theta   \right] \right| \\
& =2 \sin^2 \theta 
\label{b2t}
\end{aligned} 
\eeq
As could be expected,  the results for mode 2  are obtained from those of mode 1 by exchanging $\cos \theta \to \sin \theta$, or alternatively 
$ g_1 \to  g_2$. 
For  our  tripartite state, using Eq.\eqref{etavar}, a sufficient condition for inseparability of mode 2  is 
\beq
\gbar z >   \ln{ \sqrt{ 1 +   \left| g_1/g_2\right|^2}}   
\label{var2} 
\eeq
which again become more and more demanding as  $g_2 \to 0$, i.e. as the coupling of mode 2 become weaker.  The  bound is shown in Fig.\ref{fig_3ent} by the red curve labelled as "mode 2". 
\par
We notice that for $\cos \theta = \sin \theta  $  (i.e. $g_1=g_2$) the bounds \eqref{b1t} and \eqref{b2t} coincides, and are always smaller than  the bound  for mode  0 in Eq.\eqref{b0t}.    Therefore, a sufficient condition for genuine tripartite entanglement which holds for any  3-mode state  is that $  
 \langle[  \delta   \X_0 - ( \delta  \X_1 + \delta  \X_2)/\sqrt{2} ]^2  \rangle + \langle [\delta  \Y_0 + ( \delta  \Y_1 + \delta  \Y_2)/\sqrt{2} ]^2\rangle < 2 $, and we retrieve the VanLoock-Furusawa criterion  
formulated  in  Eq.(21) of  \cite{VanLook2003}   (where a  factor 4 in the numeric value of the bound arises from the different definitions of field quadratures). 
In the non-symmetric case,   we find more in   general  that a   sufficient criterion for genuine tripartite entanglement is 
\beq
 \langle[  \delta   \X_0 -  \delta   \X_1 \cos \theta - \delta  \X_2 \sin \theta  ]^2  \rangle + \langle [\delta  \Y_0  + \delta  \Y_1  \cos \theta +  \delta  \Y_2  \sin \theta ]^2\rangle 
< 4  \min {\left[   \cos^2 \theta , \sin^2 \theta  \right]}
\eeq
We stress that this is not the more general inequality for the  variances of nonlocal observables,  but  it is one of  the many  which can be  derived  by systematic application of the Heisenberg-like inequalities  in \eqref{PPTvar}. 
\par
When applied to our tripartite state, the criterion is satisfied provided that 
\beq
 \gbar z >    \begin{cases}  \ln{ \sqrt{ 1 +   \left| g_2/g_1\right|^2}}   
  & \text{for } |g_2| \ge  |g_1| \\
 \ln{ \sqrt{ 1 +        \left| g_1/g_2\right|^2}} 
& \text{for } |g_2| < |g_1|
\end{cases}
\eeq 
Fig.\ref{fig_3ent} summarizes the results for the tripartite entanglement of the state. Part a) of the figure shows  the results of  the test based on  the  violation of the Heisenberg-like inequalities (\ref{b0t}, \ref{b1t}, \ref{b2t}) for the mirrored obervables:   violation always occurs  for mode $\azero$,  which is never unseparable,  while 
the blue and red curves are  the boundaries \eqref{var1} and \eqref{var2} above which 
 the entanglement  of $\auno$ and  $\adue$ is verified . Therefore,  the white zone above both curves guarantees the presence of  a genuine tripartite entanglement, according to this test.  We stress that these criteria are sufficient but not necessary, therefore the shaded regions below the curves may or may not be entangled. 
\\
Parts b) and c) of the figure, conversely, shows the result of the more powerful criterion directly based on the positivity of the partial transpose, specifically on the request in Eq. \eqref{VPTsympl}
that  all the symplectic eigenvalues  of $\V^{ \mathsf{\, \scriptscriptstyle PT(A)}}$   are not below  $1$  for A-separable states.  For this  low-dimensional case, the symplectic eigenvalues  can be analytically calculated (Appendix \ref{app_sympl}). For mode 1,   the smallest symplectic eigenvalue of the partial transpose $\V^{ \mathsf{\, \scriptscriptstyle PT(1)}}$   is 
\beq
\nu_{-}^{(1)}  = 1 + 2  [\cos\theta \sinh {( \gbar z)}]^2 - \sqrt{ \left[1 + 2[ \cos \theta \sinh {( \gbar z)}]^2\right]^2 -1 }  
\label{numeno1} 
\eeq
which is always smaller than 1, unless  $\cos \theta  = 0$ i.e. unless the coupling  $g_1$ vanishes.  The result for mode 2 can be obtained  by switching $\cos \theta \to \sin \theta$:  therefore also  $\V^{ \mathsf{\, \scriptscriptstyle PT(2)}}$  always has a   symplectic eigenvalue $\nu_{-}^{(2)}  <1 $, a part from the limit case   $\sin \theta =0$, when the coupling strength $g_2$ vanishes.  
Fig.\ref{fig_3ent}b) and \ref{fig_3ent}c)  plot contour lines of  $\nu_{-}^{(1)}$ and $\nu_{-}^{(2)}$  in the parameter plane $ ( \frac{ |g_2|} {|g_1|}, \gbar z)$.  It is important to remark that   a  symplectic eigenvalue $\nu_\alpha  <1 $ of the partial transpose not only certifies the entanglement, but also provides a  measure of the bipartite entanglement, via e.g., the logarithmic negativity $ \mathit E_{\mathcal N}=- \sum_{\nu_\alpha < 1 }   \log_2  \nu_\alpha  $ \cite{Vidal2002}, with smaller   $\nu_\alpha   $ corresponding to  a larger amount of entanglement.     
\par 
In conclusion,   the  state \eqref{3state}  always exhibits a  genuine  tripartite entanglement, revealed  by the PPT criterion, which however  becomes weaker in the limit of a strong unbalance between the two coupling strength  $ |g_2/g_1| \to 0$ or $  |g_2 /g_1| \to \infty$.    In these limit situations, a larger amount of squeezing is necessary in order to detect the entanglement by means of  the less powerful criteria based on violations of Heisenber-like bounds for the variances of nonlocal observables. This state is indeed a very good example of the fact that these criteria, although  very useful and accessible, are indeed only sufficient to verify entanglement, and may fail to detect weakly entangled states. 
 
\section{Doubly pumped bulk crystal, or single-pump \HEX: 4-mode  linear coupling. }
\label{sec_Quadripartite}
An interesting feature of the 3-mode coupling generated
by the hexagonally poled NPC is that  several  independent triplets of entangled modes coexist  at   any pair of conjugate wavelengths of the fluorescence radiation. As highlighted by Ref.\cite{Jedr2018},  by  tilting the direction of propagation of the pump field inside the medium,  it is possible to reach special {\em resonance} conditions in which pairs of  triplets, originally uncoupled, degenerate into a group of 4 entangled modes. This process takes place  when the periodicity of the transverse phase modulation  of the pump  exactly matches that of the poling pattern, it involves the whole emission spectrum,  and is accompanied by a sudden enhancement of the local gain  \cite{Jedr2018}. Specifically,  by considering the  two triplets $\{ \azero, \auno,\adue \} $ and 
$\{ \azero',  \auno',\adue'\}$,  at  resonance the shared mode of each group   superimpose  to a  coupled mode  of the other, e.g. $\azero \to \auno' $ and  $\azero' \to \auno$ .  This gives rise to the  4-mode coupling  shown in the graph (b2) of  Fig.\ref{fig_schemes}, whose quantum properties were described as a  {\em Golden Ratio entanglement} \cite{Gatti2018},  because of the appearance of this particular irrational number. 
Recently,  striking  similar resonances have been  demonstrated in a doubly pumped BBO crystal  \cite{Jedr2020,Gatti2020b}. Also in this second case the resonances are reached by   tilting one of the pump modes inside the nonlinear medium, but their physical origin is  quite different,  because it involves the superposition of the Poynting vector of the pump carrier, representing the mean flux of the energy injected into the medium, with   the direction of propagation of one pump mode \cite{Gatti2020b}.  As in the previous case, at resonance each two triplets of modes degenerate into a group of 4-modes, whose coupling is  described by the scheme a2)  of  Fig.\ref{fig_schemes}. The only  difference with the NPC case is the appearance of  two distinct   coupling strengths $g_1$ and $g_2$,   controlled by  the complex amplitudes $\alpha_1, \alpha_2$ of the two pumps. 
\par 
Let us focus on a specific group of 4-entangled modes $\{ \bs,\bi, \cs,\ci \}$. Their evolution equations along the sample can be found in \cite{Gatti2018} for the NPC and in \cite{Gatti2020b} for the doubly pumped bulk crystal. They  can be derived from a longitudinal momentum operator of the form 
 \beq 
 \hat  {\cal P}   = {-i \hbar} \left( g_1 \bs^\dagger \bi^\dagger +
  g_2 \bi^\dagger    \cs^\dagger +  g_2  \ci^\dagger  \bs^\dagger  - \text{h.c} \right) ,
\label{P4A}
\eeq
where $ \bs $ and $\bi$ are   {\it  shared} modes,characterized by  two links in the graphs a2) and b2), while $\cs$ and $\ci$ are   unshared modes, with a single link. 
 We notice that these graphs  correspond to a  linear chain of nearest-neighbor interactions (we show them  as open polygons to keep continuity with the previous works \cite{Gatti2020b, Gatti2018}). 
\par
Refs.\cite{Gatti2018, Gatti2020b} showed that the z-evolution of the  four modes could be decomposed into two independent standard (i.e. 2-mode) parametric processes, mixed on an unbalanced beam-splitter.  Based on these results, we can  immediately  derive the Bloch-Messiah decomposition of state. Let us first perform local phase rotations, inessential for entanglement: 
 $ (\bs, \bi ) \to e^{ \frac{ i}{2} \arg  (g_1) } (\bs,  \bi )$ 
 and 
$ (\cs, \ci ) \to e^{ i[\arg (g_2) -\frac{ 1}{2} \arg  (g_1)] }  (\cs, \ci) $,   
which  eliminates from the problem the phases  of  $g_1$ and $g_2$ . This  physically  means that   the phase(s) of the pump beam(s) are irrelevant for the entanglement of the  state\footnote{In practice the pump phases are relevant because  they determine  the directions in phase-space where squeezing occurs}.  Thus, as in the previous section, the parameter space is spanned by the two variables 
\beq
\begin{aligned}
&  x = \frac{|g_2|} {|g_1|}  ,
\quad 
&   \rbar = \sqrt{|g_1|^2 + |g_2|^2} \,  z
\label{par} 
\end{aligned} 
\eeq
representing  the unbalance between the two processes and the total squeezing/gain available for a medium of length $z$, respectively. 
 Next we consider the transformation 
\beq
\begin{aligned} 
&\begin{pmatrix}
\bs \\
\cs\\
\bi \\
\ci 
\end{pmatrix}    = 
 \frac{1}{\sqrt{2}} 
 \begin{pmatrix}
 \cos \gamma &  0  & -\sin \gamma & 0 \\
0& \cos \gamma & 0 & -\sin \gamma \\
 \sin \gamma& 0  &\cos \gamma & 0  \\
0& \sin \gamma  &0 & \cos \gamma
\end{pmatrix} 
\begin{pmatrix}
1&  \phantom{,,}1& 0&  \phantom{,,} 0 \\
1&-1& 0&   \phantom{,,} 0 \\
 0 & \phantom{,,} 0 & 1&   \phantom{,,}1 \\
0& \phantom{,,}0& 1& -  1
\end{pmatrix}
\begin{pmatrix}
\sigmap \\
\sigmam \\
\deltap\\
\deltam
\end{pmatrix} 
& = \UB   
\begin{pmatrix}
\sigmap \\
\sigmam \\
\deltap\\
\deltam
\end{pmatrix} 
\end{aligned}
\label{UB4}
\eeq 
where $\sigmap, \ldots \deltam$ are independent bosonic modes, and the angle $\gamma$ is defined by 
\beq
\begin{aligned}
& \tg {\gamma} 
=\sqrt{\frac{\Lambda_\DD}{\Lambda_\SS} };  \qquad  & \Lambda_{\SD} (x) = \frac{\gbar}{\sqrt{x^2+1} }  \frac{\sqrt{4x^2 +1} \pm 1 }{2}
\end{aligned} 
\label{LSD} 
\eeq
Under this transformation the momentum operator  in Eq.\eqref{P4A} becomes the product of 4 independent quadratic momenta, 
\beq 
 \hat { \cal U}_B^\dagger  \, \hat {\cal P}  \,  \hat { \cal U}_B = - i \hbar   \left( 
\frac{\LS}{2} \sigmap^\dagger \sigmap^\dagger
-\frac{\LS}{2} \sigmam^\dagger \sigmam^\dagger
-\frac{\LD}{2} \deltap^\dagger \deltap^\dagger
+\frac{\LD}{2} \deltam^\dagger \deltam^\dagger - h.c.
\right) 
\eeq
where  ${ \mathcal U}_B$ is  the generator of the transformation \eqref{UB4}. 
 As a result,   the z-evolution of the 4-mode state decouples into  the product of 4 single-mode squeeze operators $\hat { {\cal S}}_j  (r) = \exp{ (\frac{r}{2}  \hat s_j^\dagger \hat s_j^\dagger -h.c.)}$,  each acting on mode  $j=1, \ldots 4$: 
\begin{align}
\hat { \cal U}_B^\dagger  e^{ \frac{i} {\hbar} \hat {\cal P} z}  \hat { \cal U}_B
=  \hat { \cal S}_1  (\LS z)  \otimes  \hat { \cal S}_2  (-\LS z ) 
 \otimes \hat { \cal S}_3  (-\LD z)  \otimes  \hat { \cal S}_4  (\LD z ) .
\label{Quad-red}
\end{align} 
where the squeeze parameters are given by Eq.\eqref{LSD}. 
\begin{figure}[ht]
\includegraphics[scale=0.55]{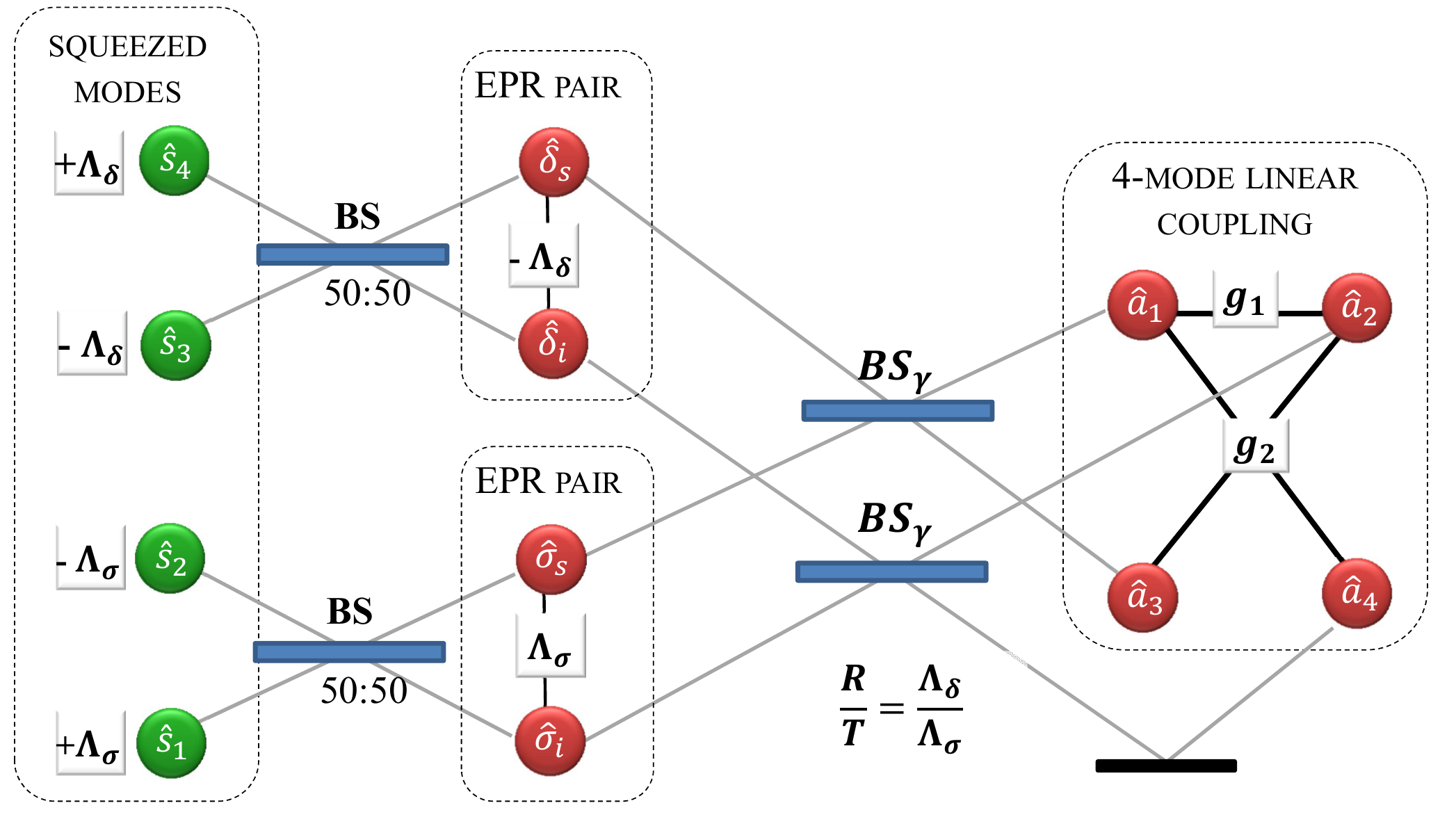}
 \caption{ Bloch-Messiah decomposition  of the 4-mode  state \eqref{P4A},   generated at resonance by a doubly pumped BBO or by a single-pump \HEX .  $\sigmap, .. \deltam$  are  independent  squeezed  modes, whose squeeze parameters $\pm \LSD z$   depend on $|g_1|$ and $|g_2|$  according to Eq.\eqref{LSD}. 
  $BS$ are   balanced beam-splitters, while    $BS_\gamma$ have  $\frac{R}{T}  = \tg^2 \gamma = \frac{\LD}{\LS}  $.   For the \HEX,  $\LS=g \Phi$, $\LD=  g/\Phi$, and  $R/T= 1/\Phi^2 $, where $\Phi$ is the Golden Ratio.  }
\label{fig_equiv4}
\end{figure}
For a vacuum input, the dynamics of the 4-mode state can be unfolded according to the scheme in Fig.\ref{fig_equiv4}:  here the 4  squeezed modes 
obtained from the action of the  squeezers in Eq.\eqref{Quad-red}
are first mixed on balanced beam-splitters to form two independent pairs of EPR modes,  labelled as $\sigma_s , \sigma_i$ and $\delta_s,\delta_i$ -where " s"  and  "i" stand for signal and idler-  with squeezing parameters $\LS z $ and $-\LD z$, respectively.  Next the EPR modes are pairwise mixed (signal with signal and idler with idler) on two identical beam splitters, whose splitting ratio is linked to the squeeze eigenvalues by 
\beq
\frac{R}{T}= \frac{\LD}{\LS}, 
\label{cond_lin}
\eeq
 Indeed, this is the  condition which  ensures that modes  $\cs$ and $\ci$ are uncoupled, and that the graph of the coupling reduces to  a linear chain. 
\par 
We notice that for  $x=1$ $\Lambda_\SS  \to { \gbar \over \sqrt{2}}  \Phi $ and $\Lambda_\DD \to  { \gbar \over \sqrt{2}}    {1 \over \Phi}$, where $\Phi = \half (1+\sqrt{5})$ is the Golden Ratio.  Thus for the single-pump \HEX , the squeeze parameters are $ g \Phi  $ and $g/\Phi $, and the "Golden Ratio"  squeezing  or gain enhancement  described in \cite{Jedr2018,Gatti2018} takes place.
 For the doubly-pumped BBO, one eigenvalue is always smaller than $\gbar$:  
$ \LD/\gbar \in (0,1)$ but  the other  is always larger:  $ {\LS}/\gbar \in (1, \frac{2}{\sqrt 3})$, with the maximum value $  2/\sqrt{3} = 1.15..$ occurring  when pump 2  is twice as intense as pump 1. Thus, at difference with the 3-mode case, the doubly-pumped scheme at resonance exhibits   a true increase of the squeezing/gain level of certain modes, with respect to the use of a single-pump with  the same energy . 
 \par
\subsection{Quadripartite entanglement of the  linear state.} 
As for the tripartite state,  the 4-mode entanglement will be characterized  both by using   criteria based on variances of nonlocal observables [Eq.\eqref{PPTvar}], and by means of  the  symplectic eigenvalues of the partial transpose [Eq. \eqref{VPTsympl}]. 
\par
Inversion of the matrix \eqref{UB4} permits to identify the nonlocal variables with sub-shot noise fluctuations. Taking into account the sign of the squeezing parameters in Eq.\eqref{Quad-red},    
 these are: 
\bsub
\begin{align}
\hat \eta ( \vd_\SS^\prime) & 
= \frac{1}{\sqrt 2} \left[  \cos \gamma \left(\Y_1 +\Y_2 \right)  + \sin \gamma \left(\Y_3 + \Y_4\right)  \right]  = e^{-\LS z}  \hat Y_{s_1}^\mathrm{in}
\label{etasigmap} \\
\hat \eta ( \vd_\SS) & 
= \frac{1}{\sqrt 2} \left[  \cos \gamma \left(\X_1 -\X_2 \right)  + \sin \gamma \left(\X_3 - \X_4\right)  \right]  = e^{-\LS z}  \hat X_{s_2}^\mathrm{in}
\label{etasigmam} \\
\hat \eta ( \vd_\DD) & 
= \frac{1}{\sqrt 2} \left[  -\sin \gamma \left(\X_1 +\X_2 \right)  + \cos \gamma \left(\X_3 + \X_4\right)  \right]  = e^{-\LD z}  \hat X_{s_3}^\mathrm{in}
\label{etadeltap} \\
\hat \eta ( \vd_\DD^\prime) & 
= \frac{1}{\sqrt 2} \left[  -\sin \gamma \left(\Y_1 -\Y_2 \right)  + \cos \gamma \left(\Y_3 - \Y_4\right)  \right]  = e^{-\LD z}  \hat Y_{s_4}^\mathrm{in}
\label{etadeltam} 
\end{align}
\label{eta4}
\esub
By construction, the variables commute pairwise $\left[ \hat \eta ( \vd_\alpha) , \hat \eta ( \vd_\beta^\prime)  \right ] = 0 $ and  their  variances asymptotically vanish in the limit of large squeezing $\gbar z >>1$
\begin{align}
\llangle \delta \hat \eta^2 (\vd_\alpha) \rrangle = \llangle \delta \hat \eta^2 (\vd_\alpha^\prime) \rrangle  
=e^{-2\Lambda_\alpha z} , \qquad (\alpha=\sigma,\delta)
\label{vareta4} 
\end{align} 
Conversely, if we consider  states that are separable with respect to  a given partition $ \{A\} \times  \{B\} $ of the set of four modes, the  variances  of the observables \eqref{eta4} are constrained by the  four bounds that arise from the  nonzero commutators of the mirrored variables. Precisely, by using Eq. \eqref{PPTvar}, these bounds are 
\bsub
\label{bound4}

\begin{align}
\llangle \delta \hat \eta^2 (\vd_\alpha) \rrangle + \llangle \delta \hat \eta^2 (\vd_\alpha^\prime) \rrangle 
 & \ge \left| \left[ \eta ( \PT_A \vd_\alpha) ,  \eta (\PT_A \vd_\alpha^\prime ) \right] \right| 
\qquad (\alpha=\sigma,\delta)
\label{sumvar} \\
2\sqrt{ \llangle \delta \hat \eta^2 (\vd_\alpha) \rrangle \,  \langle \delta \hat \eta^2 (\vd_\beta^\prime) \rangle}  
&  \ge \left| \left[ \eta ( \PT_A \vd_\alpha) ,  \eta (\PT_A \vd_\beta^\prime ) \right] \right| 
\qquad (\alpha \ne  \beta = \sigma,\delta)
\label{prodvar} 
\end{align} 
\esub
where  $\PT_A$ is the mirror matrix defined by Eq. \eqref{Mirror}. Notice that  for  $\alpha \ne \beta$ [Eq. \eqref{prodvar}]  we directly wrote   the criterion  for  the  product  of variances   because in this case    it is   more stringent  than the one  for the sum.  Violation of any of these bounds is sufficient to verify that the state is not A-separable. 
\par 
For each of the possible partitions of the system, we checked the four constraints, and we chose among them the one that is violated first in terms of the total gain (see table \ref{table_bound4} in Appendix \ref{app_bounds}). The  seven distinct partitions of four modes are  listed below: 
\renewcommand{\arraystretch}{1.4}
\setlength{\tabcolsep}{4pt}
\begin{center}
\begin{tabular}{ | c | c |  c  | c |c|c|c| } 
\hline
$P_1$                                   & $P_2$                                        & $P_3$                       & $P_4$ &     $P_{12}$           & $P_{13}$        & $P_{14}$      \\
  \hline
$\{1 \}  \times \{2,3,4 \}$  & $\{2\}  \times \{1,3,4 \}$ & $\{3\}  \times \{1,2,4 \}$ &  $\{4\}  \times \{1,2,3 \}$ & $ \{1,2 \}  \times \{3,4 \}$	&  $ \{1,3 \}  \times \{2,4 \}$	&  $ \{1,4 \}  \times \{2,3 \}$	\\
\hline
\end{tabular} 
\end{center} 
\begin{figure}
\includegraphics[scale=0.68]{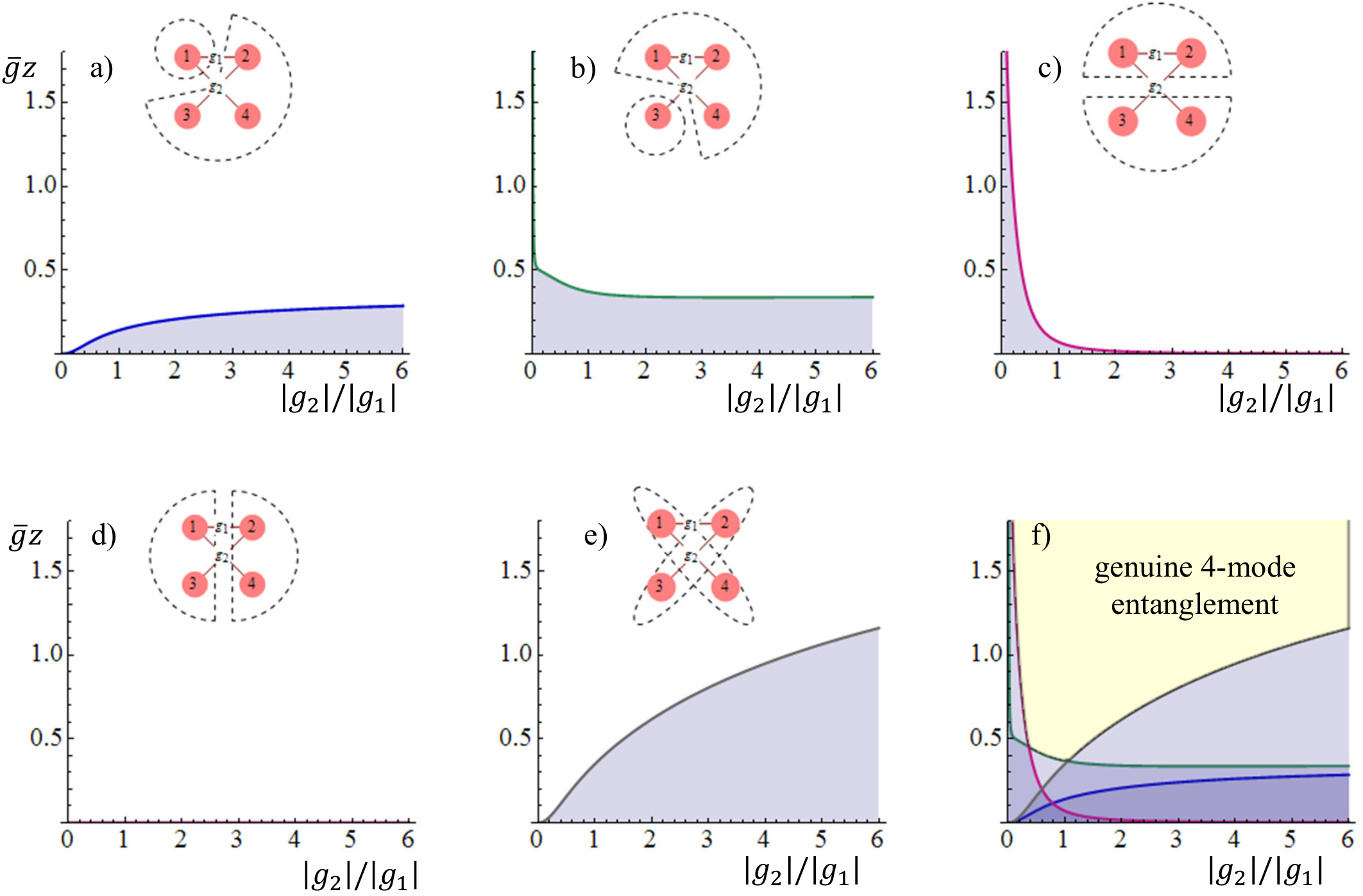}
 \caption{Entanglement of the 4-mode linear  state \eqref{P4A}, according to sufficient criteria    based on  the violation of the  Heisenberg-like inequalities \eqref{bound4}. Plots  a) to e)  are the results  for partitions $P_1$,  $P_3$, $P_{12}$, $P_{13}$ and $P_{14}$, respectively, where inseparability  
 is guaranteed in the region above the curve. The plots for  $P_2$ and $P_4$ are identical to those for $P_1$ and $P_3$, respectively.  f) summarizes all the bounds, with a genuine 4-party entanglement being verified in the yellow region above all curves. 
}
\label{fig_bounds4}
\end{figure} 
The panels  from  a)  to e) of Fig.\ref{fig_bounds4} show the results obtained for each  bipartition. The curves plot   the lowest one of the constraints  \eqref{bound4},  so that in the  white region above each curve all the Heisenberg-like inequalities \eqref{bound4}  are violated, which  guarantees that the state is inseparable with respect to the given partition.  We remind that the criterion is only sufficient, so that the shaded regions below each curve may or may not be entangled.
For reasons of symmetry,  the results for $P_2$ and $P_4$  are identical to those for $P_1$ and $P_3$, respectively, and  therefore are not shown.   Panel f) summarizes all the bounds, where the  yellow region above all  curves  guarantees the presence of  a genuine quadripartite entanglement, according to this test.
Notice that  violation always occur for partition $P_{13}$  [panel d)], implying that  for any finite gain $\gbar z \ne 0$  the state is never fully separable. This is  a straightforward consequence of the existence   in the  subspace  $\{\bs,\cs\}$ of  linear combinations of modes, namely  the EPR signal   modes $\sigma_{s}$  and $\delta_{s}$ in figure \ref{fig_equiv4},  that are maximally entangled to
their EPR partners  $\sigma_{i}$  and $\delta_{i}$ living  in the  complementary  subspace  $\{\bi,\ci\}$. \\
\begin{figure}
\includegraphics[scale=0.69]{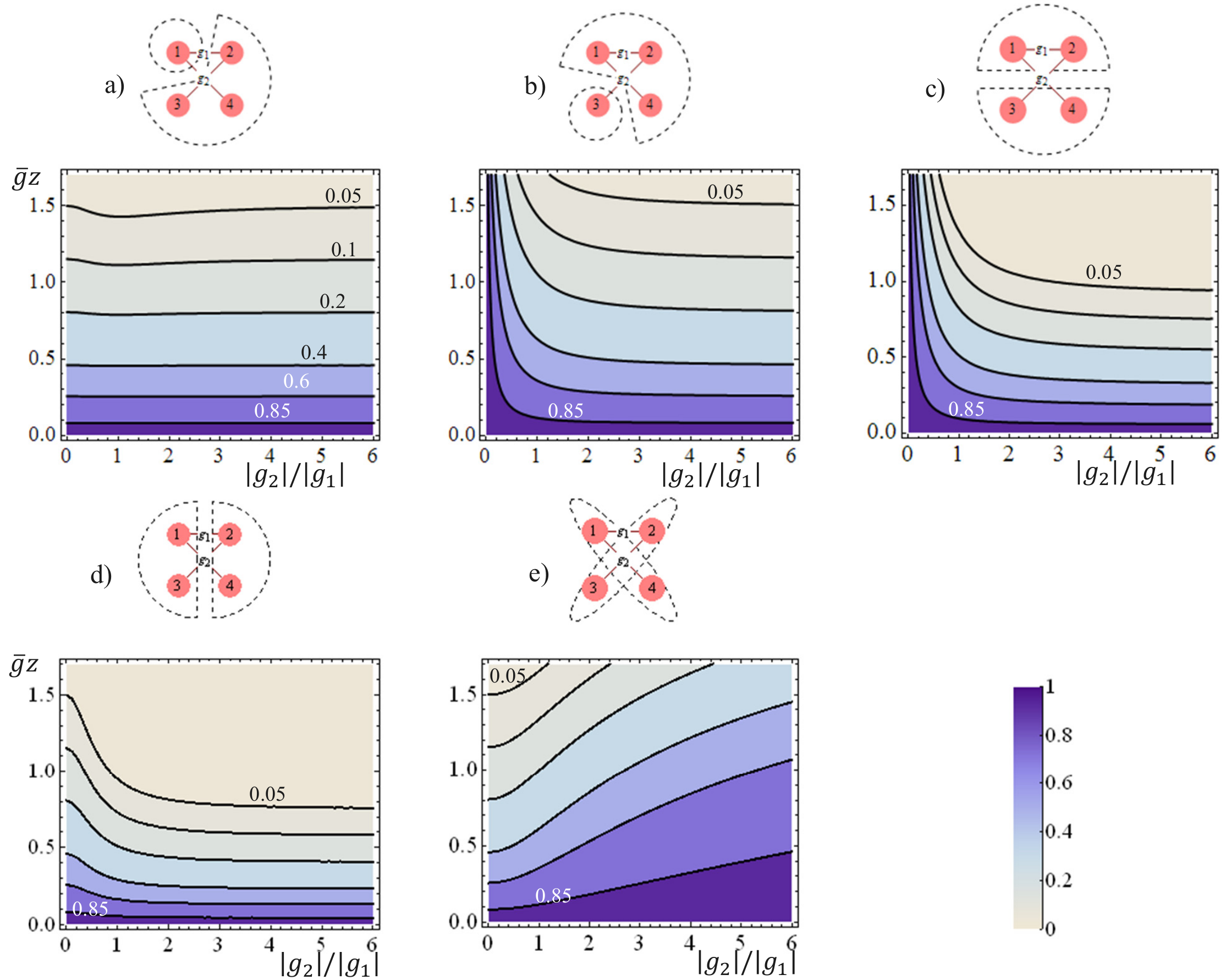}
 \caption{Entanglement of the 4-mode linear  state \eqref{P4A}, according to the PPT criterion. 
a) to e)  are
contour plots   of the  symplectic eigenvalues   $\nu_{\alpha} <1 $   of  $V^{\PTA A } $
 listed in Table \ref{table_sympl4},  for each partition.   For $P_1$,  $P_3$ and $P_{14}$ there is a single  eigenvalue  $\nu_\alpha < 1$, shown in a), b) and e) ( the results  for   $P_2$ and $P_4$ are identical to those for $P_1 $and $P_3$, respectively, and are not shown),      For
$P_{12}$ and $P_{13}$,  panels  c) and d) show the product of   the two eigenvalues below 1.   
}
\label{fig_symplectic4}
\end{figure} 
Fig.\ref{fig_symplectic4} shows instead the results of the more powerful test based on the  PPT criterion, namely on the  request [Eq. \eqref{VPTsympl}]
that    for A-separable states all the symplectic eigenvalues of $\V^{ \mathsf{\, \scriptscriptstyle PT(A)}}$  satisfy  $\nu_\alpha \ge 1$. For  each of the seven bipartitions, the symplectic spectra have been analytically calculated, and are reported in table   \ref{table_sympl4} of  Appendix \ref{app_sympl}.  Fig.\ref{fig_symplectic4}  shows  contour plots,  in the parameter plane $(|g_2/g_1|, \gbar z) $,   of   the  product $\prod_{\nu_\alpha <1} \nu_\alpha $.  
We remind that this quantity provides a quantitative assessment of the bipartite entanglement, via the logarithmic negativity $ \mathit{ E}_{\mathcal N} =- \log_2 \left( \prod_{\nu_\alpha <1} \nu_\alpha \right) $ \cite{Vidal2002}. \\
As already noticed for the 3-mode state,  the PPT criterion reveals the presence of  entanglement  in regions of the parameter space where 
the Heisenberg-like criteria \eqref{bound4} fail to detect it. Actually,  according to the results in Fig.\ref{fig_symplectic4}, the four-mode linear state  exhibits  a genuine 4-party entanglement for any finite value of the gain $\gbar z$, unless  trivially one of the two couplings vanishes,  $g_1 = 0$ or $g_2 =0$ 
  In particular, for the single-pump NPC, where $g_2=g_1 = \sqrt 2 \gbar$,  the 4-mode entanglement is always a genuine one. \\
Indeed the analysis of the symplectic spectra provides a quantitative assessment of what could be intuitively expected for a linear chain of nearest neighbor interactions. Assuming a given amount of  the total gain $\gbar z >0$, 
  the shared modes 1 and 2,  characterized by  two links of strength $g_1$ and $g_2$ with the other modes, are never separable [panel a)], and their level of entanglement remains  fairly constant with the ratio $ |g_2/g_1| $. 
Conversely,  the entanglement of modes 3 and 4, which have the single link $g_2$, is  weaker and vanishes for   $g_2 \to 0$ [panel b)].  Concerning the 2 $\times$ 2 bipartitions, we notice that $P_{12}$  [panel c)] has  a double link of  strength $g_2$: therefore,   it   may  become separable for  $g_2 = 0$, but has a quite high level of entanglement when  $|g_2|  > |g_1|$.  In comparison, 
$P_{14} $ [panel e)],  which becomes separable for $g_1 \to 0$, has a weaker entanglement because it has 
a single link.  
The "signal-idler" 
 partition  $P_{13}$ [panel d)] is clearly the most entangled one: it is never separable, because its three  links of different strengths cannot simultaneously  vanish for  $\gbar z >0$.  In this case,  the 
 symplectic eigenvalues $\nu_\alpha < 1$  of the partial transpose have  the  particularly simple expressions  $e^{-2 \LD z}$ and $e^{-2 \LD z}$ [table \ref{table_sympl4}],  where $\LS z $ and $\LD z $ are the squeeze parameters of the Bloch-Messiah decomposition of the state, given by  Eq. \eqref{LSD}. Using this equation, after some simple calculations, 
the logarithmic negativity  turns up 
$ \mathit E_{\mathcal N}  
 =  \frac{2 }{\ln 2} \sqrt{ 4 |g_2|^2 + |g_1|^2}$ . Thus, for 
$g_2=0$  $ \mathit E_{\mathcal N} \to  \frac{2 }{\ln 2} |g_1|  z $, which is the logarithmic negativity of a single  EPR state with gain $|g_1| z $ 
, while  for  $g_1=0  $,  $ \mathit E_{\mathcal N} \to  \frac{4 }{\ln 2} |g_2| z $, corresponding to the entanglement of two independent EPR states. 
\section{ Doubly pumped NPC: 4-mode square coupling }
\label{sec_Quadrisquare}
The last example that we consider is that of  doubly pumped   \HEX , schematically shown by  Fig.\ref{fig_schemes}c: here  the intensity pattern created by the interference of the two pump modes  superimposes to the static  pattern of the periodic poling.  The four possible combinations of the pump modes with the grating vectors $\vec G_1$ and $\vec G_2$ drive four  concurrent  processes  which create four families of intersecting cones \cite{Gatti2020,Brambilla2019}. A specially important configuration is reached at spatial resonance, when the periodicity of the pump pattern matches that of the poling pattern, and two of the four processes degenerate into a single one, whose parametric gain is controlled by the coherent sum of the two pump amplitudes $\ |\alpha_1 + \alpha_2| $ \cite{Gatti2020, Brambilla2019}.  This gives rise to   the four-mode coupling shown in Fig.\ref{fig_schemes}c2) in proper groups of spatio-temporal light modes. Remarkably, this coupling   is topologically different from the one of the single-pump NPC, because it is a closed square chain of nearest-neighbor interactions. 

\par
\begin{figure}[hb]
\includegraphics[scale=0.55]{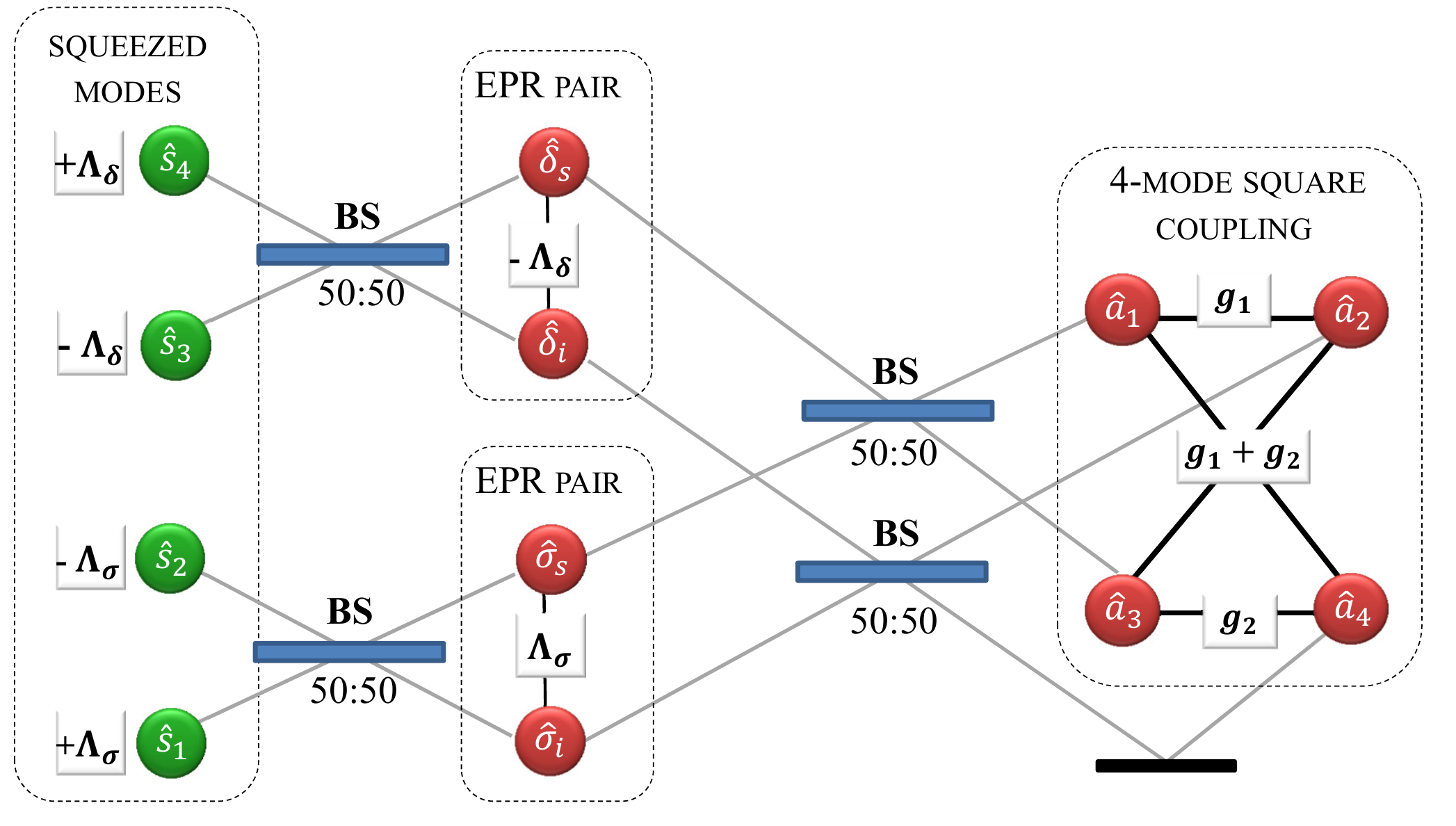}
 \caption{ Bloch-Messiah decomposition  of the 4-mode  state \eqref{P4sq},   generated at resonance by a  \HEX pumped by two pumps of equal intensity $|g_1|=|g_2|$.  Modes $\sigmap, .. \deltam$  have squeeze parameters $\pm \LSD z$ which depend on the pump phase-difference  $\phimeno$  as described by Eq.\eqref{LSDsq}.
  $BS$ are   balanced beam-splitters. 
}
\label{fig_equiv4sq}
\end{figure}

 
 As detailed in Ref.\cite{Gatti2020}, the generator of the $z-$evolution along the sample of each quadruplet of coupled modes is the longitudinal momentum 
 \beq 
 \hat  {\cal P}   = {-i \hbar} \left( g_1 \bs^\dagger \bi^\dagger +
 (g_1 + g_2 ) \bi^\dagger    \cs^\dagger + g_2 \cs^\dagger \ci ^\dagger +   (g_1 + g_2)   \ci^\dagger  \bs^\dagger  - \text{h.c} \right) . 
\label{P4sq}
\eeq
We notice that in this case all modes are   {shared}  between two of the three processes,  and are characterized by  two links in the graph of Fig. \ref{fig_equiv4sq}.   We also notice that the central links  have strengths $g_1 + g_2$  controlled by the phases of $g_1$ and $g_2$, i.e. by the phases of  the pump modes, which  represents the specific feature of this configuration,  absent in the case of  the doubly pumped bulk crystal where the  phases were irrelevant. 
\par 
In  order to emphasize this  feature (and also for reasons for brevity),  in the following  we limit our analysis to the {\em symmetric} case, in which  the pumps have equal intensities  and
\beq
|g_1| =|g_2|
\eeq
 By  performing  the local phase rotations 
 $ (\bs, \bi ) \to e^{ \frac{ i}{2} \arg  (g_1) } (\bs,  \bi )$ 
 and 
$ (\cs, \ci ) \to - e^{  \frac{ i}{2} \arg  (g_2)] }  (\cs, \ci) $,   
the momentum operator reduces to 
$   \hat  {\cal P}   = {-i \hbar} \frac{\gbar}{\sqrt 2}  \left[ \bs^\dagger \bi^\dagger +   \cs^\dagger \ci ^\dagger  + 2 \cos \phi_- (  \bi^\dagger    \cs^\dagger +     \ci^\dagger  \bs^\dagger)  - \text{h.c} \right] $. The 
 parameter space is  spanned by the two variables 
\beq
\begin{aligned}
&  \phimeno =    \frac {\arg  (g_1)  -\arg  (g_2)}{2}
\quad 
&   \rbar = \sqrt{|g_1|^2 + |g_2|^2} \,  z
\label{par} 
\end{aligned} 
\eeq
representing   the phase shift between the two processes (physically, $\phimeno$ is the offset between the  transverse modulations of the pump and of the poling pattern \cite{Gatti2020})  and the total squeezing/gain available for a medium of length $z$. 
\par
The Bloch-Messiah decomposition of state is  shown in Fig.\ref{fig_equiv4sq}; it is obtained by means of the transformation 
\beq
\begin{aligned} 
&\begin{pmatrix}
\bs \\
\cs\\
\bi \\
\ci 
\end{pmatrix}    = 
 \frac{1}{2} 
\begin{pmatrix}
1&  \phantom{,,}1&  \phantom{,,}1&  \phantom{,,} 1 \\
1&-1& 1&   - 1 \\
 1 & \phantom{,,} 1 & -1&   -1 \\
1& -1& -1&  1
\end{pmatrix}
\begin{pmatrix}
\sigmap \\
\sigmam \\
\deltap\\
\deltam
\end{pmatrix} 
& = \UB   
\begin{pmatrix}
\sigmap \\
\sigmam \\
\deltap\\
\deltam
\end{pmatrix} 
\end{aligned}
\label{UB4sq}
\eeq 
where  the squeezed modes $\sigmap, \sigmam, \deltap,\deltam$ have squeeze parameters $+\LS z ,-\LS z ,-\LD z , +\LD z$ in order, with 
\beq
\Lambda_{\SD}= \frac{\gbar } {\sqrt 2} \left( 2 \cos \phimeno \pm 1  \right)  =  |g_1 + g_2| \pm |g_1|
\label{LSDsq}
\eeq 
We notice that the  decomposition is  very similar to the one in Fig.\ref{fig_equiv4}, but the condition \eqref{cond_lin} is released, with the 
 $BS_{\gamma}$ replaced by  balanced  beam-splitters, having  
$R/T=\tg^2 \gamma =1$. 
\subsection{ Quadripartite entanglement of the  square state}
\begin{figure}{h}
\includegraphics[scale=0.7]{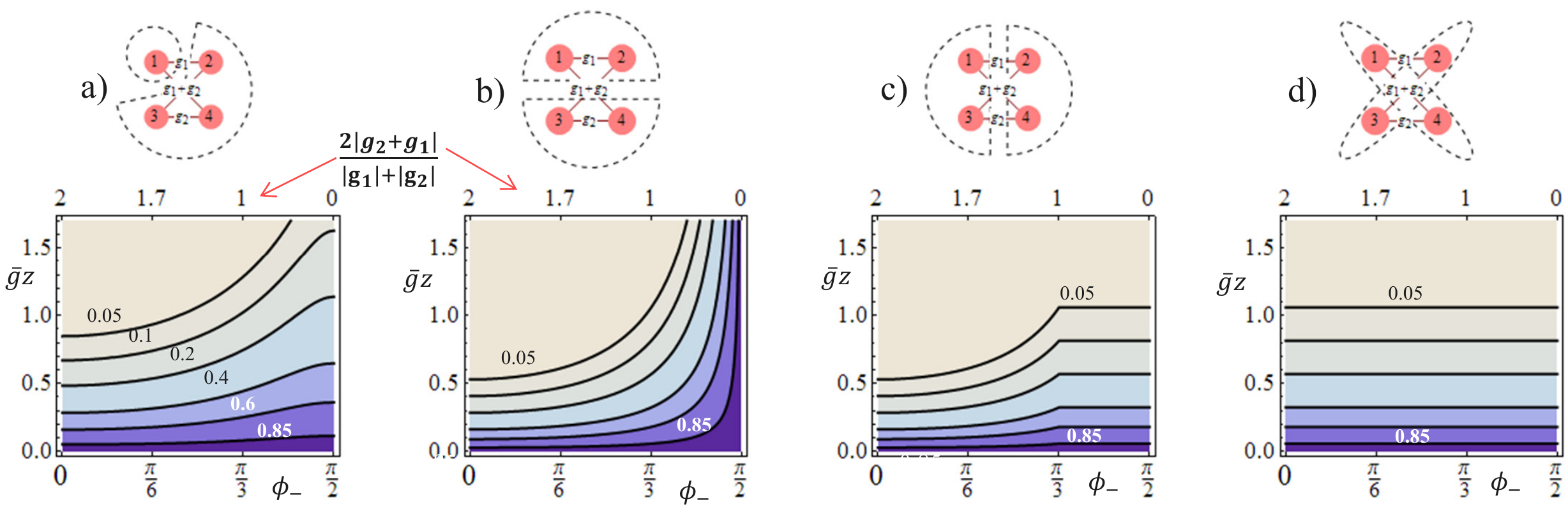}
 \caption{Entanglement of the 4-mode square  state \eqref{P4sq}, according to the PPT criterion. 
a) to e)  are
contour plots   of the product of the   symplectic eigenvalues   $\nu_{\alpha}  \le 1 $   of  $V^{\PTA A } $
 listed in Table \ref{table_sympl4sq},  for each partition of the system. a) shows the results   for the 1-mode partitions $P_j$, which are all are all identical;  b),c) and d) are the results for the remaining partitions.  The   quadripartite entanglement is always genuine, except at  the  point $\phi_-={\pi \over 2}$, where  $g_1+g_2=0$  and $P_{12}$ in b) becomes separable.
}
\label{fig_symplectic4sq}
\end{figure}

 For the state generated by  the square coupling  \eqref{P4sq} we omit the long but simple analysis based  on the  variances of nonlocal observables, which can be  performed along the guidelines of the previous sections, by inverting the transformation \eqref{UB4sq} (some caution should be taken  with the sign of $\LD$).  As in the other examples,  this kind of analysis does not actually  add anything to the results based on the symplectic eigenvalues of the partial transpose matrix $\V^{\PTA A}$, which for this simple state can be analytically performed and are   presented in Fig.\ref{fig_symplectic4sq}.
\par
 Before discussing these results, let us consider the following points: 
 According to our definitions,   $\LD$ in Eq.\eqref{LSDsq}  is positive for $\phimeno < \frac{\pi}{3}$ and negative for 
$\phimeno> \frac{\pi}{3}$. The  point   $\phimeno= \frac{\pi}{3}$  ($\!\mod \pi $)\footnote{We limit to $\phimeno \in (0,{\pi \over 2})$,  which corresponds to a phase difference between the two pumps $\in (0,\pi)$. 
}  at which $\LD=0$ and  modes $\deltap$ and $\deltam $  are not squeezed   is characterized by the fact  that all the  intermodal links  have the same strength: $|g_1+g_2| = |g_1| = |g_2|$. At this point,  the 4-mode state corresponds to a balanced mixture of an EPR state  with a vacuum or coherent state, as discussed in \cite{Gatti2020}.   The other important point is  $\phimeno =\frac{\pi}{2}$  at which  $g_1+g_2=0$, and the two central links vanish, so that the 4-mode coupling degenerates into the product of two independent 2-mode couplings.  At this point  $\LD =-\LS$,  and,  according to the  Bloch-Messiah decomposition in Fig. \ref{fig_equiv4sq},  the state is  the mixture of two identical EPR states on a balanced beam-splitter, which again results into two independent EPR pairs. Correspondingly, we expect that the quadripartite entanglement of the state ceases to be a genuine one. 
\par 
These considerations are well reflected by the analysis of the entanglement  based on the  PPT criterion.  Table \ref{table_sympl4sq} reports  our results for  the symplectic spectrum of the partial transpose matrix $\V^{\PTA A}$, and shows the conditions for having symplectic eigenvalues $\nu_\alpha < 1$ which guarantee the inseparability with respect to each given partition. 
 Fig.\ref{fig_symplectic4sq} shows countour plots of the product $\prod_{\nu_\alpha <1} \nu_\alpha$  in the parameter plane $(\phimeno, \gbar z)$  for each partition of the system.  Panel a) shows the results for the 1-mode partitions, which in this case are all identical, because of the symmetry of state. These partitions are never separable, because each mode has one link  of strength $ |g_1|=|g_2|$ that never vanishes, while  the strength of the other   $|g_1+ g_2|= 2 |g_1| \cos \phimeno$ decreases from  $\phimeno=0$ to $\phimeno=  {\pi \over 2}$, and this behavior is reflected by the entanglement of the partition.  $P_{12} = \{1,2 \}  \times  \{3,4 \} $ is the only partition which becomes separable at $\phimeno =  {\pi \over 2} $, where its double link $g_1+g_2$ vanishes.  The signal-idler partition $P_{13}$ is always entangled, and so does the partition $\{1,4 \}  \times  \{2,3 \} $: in this last case the amount of entanglement remains constant with the phase-difference $\phimeno$, because the strengths of its two links are $|g_1|=|g_2|$. 
\section{Discussion and Conclusions} 
There are two kind of conclusions which can be drawn from this analysis. 
\par
From the point of view of physics, it has been shown that a  genuine multipartite  entanglement among separate spatial modes of the electromagnetic field can be generated by spatially engineering the effective nonlinearity of parametric sources. In this work, this is achieved by a simple transverse modulation of the pump beam driving the process and/or of the nonlinear response in a NPC.  
The quantum states generated in this way are rather simple, since the entanglement is limited to sets of three or four modes; as such, they cannot compete at the present stage of analysis with the multipartite states generated by more powerful schemes based on  on tailoring the spectral entanglement of  frequency combs   \cite{Physer2011,Zhu2021, Roslund2014,Cai2017}. However more complex setting may be envisaged, involving more spatial modes of the pump beam or more reciprocal vectors of the nonlinear pattern of the NPC. For example, one may think of driving a parametric process by means of N plane-wave pumps, whose wavevectors are symmetrically tilted  over a conical surface, similarly to the scheme proposed in Ref. \cite{Daems2010}. Preliminary investigations have shown \cite{Gatti2021} that there exist realistic configurations (in terms of the choice of the material, of the type of interaction, of the frequencies involved etc.) in which  N modes of the down-converted radiations  may become coupled one to each other, according to a closed chain of nearest neighbor interactions, realizing thus a higher dimensional version of the square 4-mode entanglement described in Sec.\ref{sec_Quadrisquare}.  In the same setup, by slightly changing the parameters  it can be also achieved a situation in which  there is a single central mode  of the down-converted light  coupled to each of  N spatially separated modes \cite{Gatti2021}, thus realizing the multipartite entanglement analysed in \cite{Daems2010}. \\
It should be also noticed that even in the simpler setup considered in this paper (N=2, or equivalently,   two reciprocal vectors of the nonlinear grating)  there are actually  infinite sets of 
 triplets  (or quadruplets) of entangled modes which coexist independently in the same beam. These are  distinguished by their spatial coordinates in one transverse direction, say $y$,  as e.g. shown by Figs. 6 and 7 in  \cite{Gatti2018}. Strategies for entangling independent groups of modes  in a higher dimensional  state could be envisaged; for example by employing  in the spatial domain techniques analogue to those used in the temporal domain  to generate the XEPR state in Ref.\cite{Furusawa2013},  with  the temporal delay being  replaced by a spatial shift. \\
We hope that the proof-of-principle analysis performed in this work may stimulate experimental work in this sense.
\par 
The second conclusion is more on the formal side and concerns the effectiveness of the various strategies for detecting entanglement. We formulated general bounds  [Eq. \eqref{PPTvar}] which must be respected by  states that are separable with respect to a given partition of the system. As previously mentioned, these inequalities are not fundamentally new compared to similar ones that can be found in the existing literature \cite{VanLook2003, Teh2014,Toscano2015}, but have the merit of a very general and expressive form, in which  the variances of nonlocal observables are bounded by the commutators of  the corresponding "mirrored" observables. Moreover,   for a given Gaussian state, we suggested a simple method for identifying the nonlocal observables that are most likely to  certify  its  entanglement, based on its Bloch-Messiah decomposition.   The states examined in this work, thanks to their relative simplicity,  gave us the possibility of verifying to what extend these criteria are efficient in detecting  the entanglement.   By performing analytical calculations of the  various symplectric spectra involved,    we have shown  that, although weaker than the  PPT criteria which require  the whole covariance matrix, the more experimentally accessible  criteria based on  the Heisenberg-like inequalities \eqref{PPTvar} capture the essential structure of the entanglement of the state (but, clearly,  they are unable to  give any quantitative assessment). 
\newpage
\appendix 
\section{Symplectic spectra}
\label{app_sympl}
\subsection{Williamson normal form and symplectic eigenvalues } 
\label{app_Williamson}
According to a theorem due to Williamson \cite{Williamson1937}, any real, positive-definite, symmetric $ 2N \times 2N$ matrix can be brought to a diagonal form 
by means of a linear basis change that preserves the symplectic form $\Sy$. When applied to the  covariance matrix, it implies that any covariance matrix can be written as 
$\V = \vect{S} \V^\mathrm{in}  \vect{S}^\intercal$, where $\V^\mathrm{in}= \mathrm{diag}\{\nu_1,\nu_2 \ldots \nu_N, \nu_1, \nu_2 \ldots \nu_N\}$, 
and 
$\vect{S}$ is a  {  symplectic} transformation , i.e. a  2Nx2N matrix  that  preserves the fundamental commutation relation $\vect{S}^\intercal \Sy \vect{S} = \Sy$.  The set of  N  positive numbers $\vect{\nu} = \{\nu_1,\nu_2 \ldots \nu_N \} $ is called the symplectic spectrum of $\vect{V}$, and can be obtained by calculating  the eigenvalues of $  i \Sy \V $ which are $ \pm \nu_1, \pm \nu_2 \ldots \pm \nu_N$.  Then, the inequality \eqref{positivity} implies that for a  legitimate covariance matrix  $\nu_ \alpha \ge 1$ , $ \forall  \alpha=1,\ldots,N$, which physically corresponds to $\V^\mathrm{in}$ being the covariance matrix of N independent thermal states. For pure Gaussian states,   $\nu_ \alpha = 1$ , $ \forall  \alpha=1,\ldots,N$. 
\par
Considering a partition  $ \{ A\} \times \{ B \}$ of the system, the  partial transpose matrix  $\V^{ \mathsf{\, \scriptscriptstyle PT(A)}}$  is still real, positive definite and symmetric. Therefore 
the Williamson theorem can  applied  to it,  and  the symplectic spectrum  calculated from the eigenvalues of $  i \Sy \V^{ \mathsf{\, \scriptscriptstyle PT(A)}} $.  In this case,  however,   nothing ensures that  $\nu_\alpha \ge 1$: only for A-separable states, 
$\V^{ \mathsf{\, \scriptscriptstyle PT(A)}}$  is a legitimate covariance matrix,  and  its symplectic spectrum must satisfy   $\nu_\alpha  \ge 1$ , $ \forall \alpha=1,\ldots,N$. Conversely,  the appearance of an eigenvalue  $\nu_\alpha <1$ in the symplectic spectrum of the partial transpose is sufficient to demonstrate that the state is not  A-separable. 
 
\subsection{Calculation of  the symplectic spectra of the partial transpose}
\begin{figure}
\includegraphics[scale=0.67]{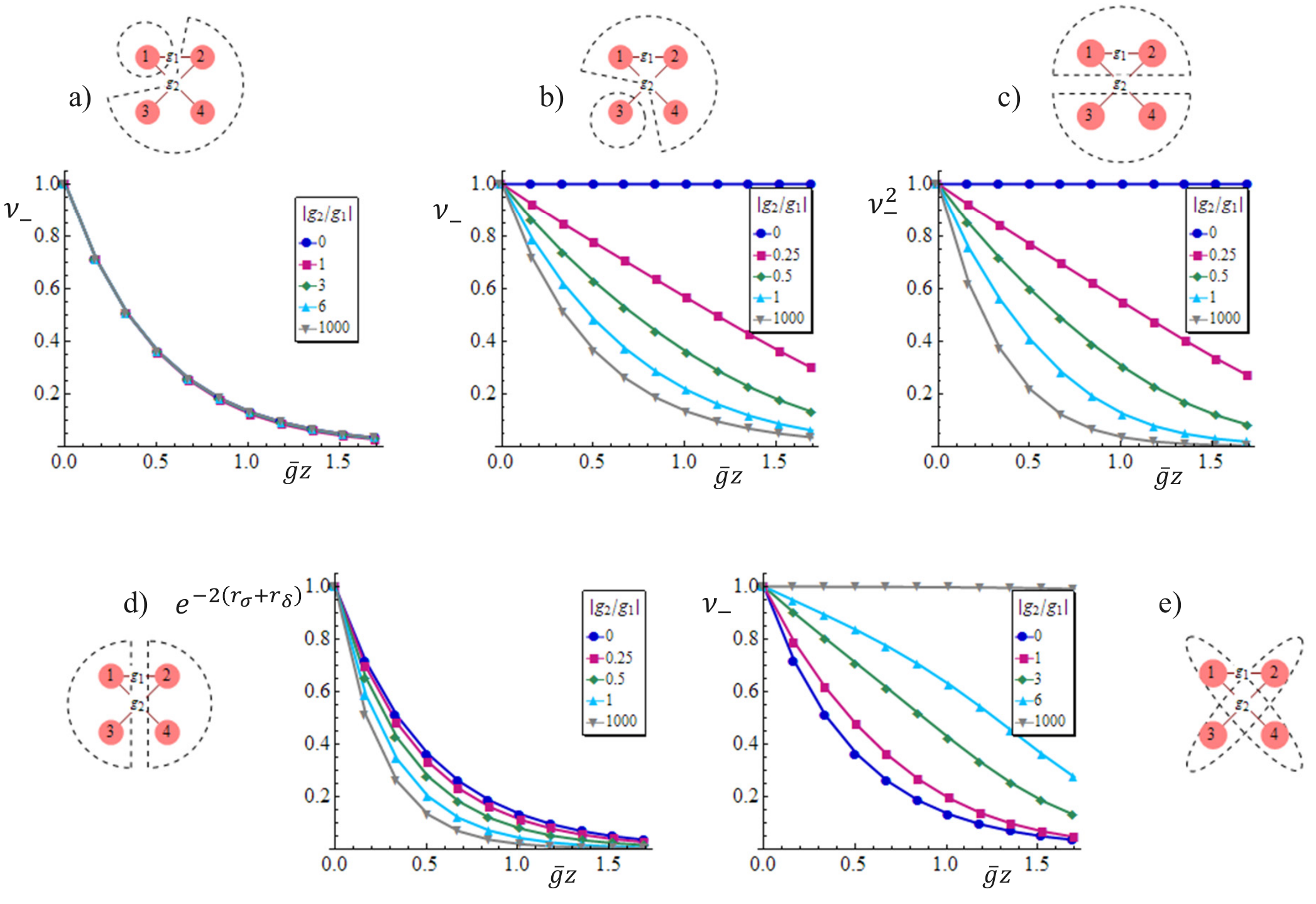}
 \caption{ {\bf Four-mode linear coupling}: symplectic eigenvalues   $\nu_{\alpha} \le 1 $   of  $V^{\PTA A }$
 listed in table \ref{table_sympl4}.     a) to e)  show
$\prod_{\nu_\alpha <1} \nu_\alpha $ for each partition.  The plots for  partitions $P_2$ and $P_4$ are identical to those for $P_1 $and $P_3$ in panels a) and b), respectively. 
}
\label{fig_list4}
\end{figure}

For each state considered  in the main text,  its  Bloch Messiah reduction was exploited in order to calculate the covariance 
matrix $\V$ associated with the state, following  the procedure described  in Sec.\ref{sec_BlochMessiah},  see  in particular  equation \eqref{Vcalc}.  For each  partition $\{A\}$$\times$$ \{ B \}$ of the system, 
the partial transpose matrix was then calculated as $ \V^{ \mathsf{\, \scriptscriptstyle PT(A)}} = \PT_A \V \PT_A $, where  $\PT_A $  is  the mirror transformation \eqref{Mirror}. Finally, the characteristic polynomial associated to the matrix   $  i \Sy \V^{\PTA A} $ was analysed, and  its eigenvalues calculated 
with the help of Wolfran Mathematica \cite{Mathematica}.   
\\
Results  for the 3-mode state \eqref{3state} are shown in table \ref{table_sympl3}. The results  for the 4-mode linear state \eqref{P4A}  are provided by table \ref{table_sympl4}, while figure \ref{fig_list4} plots  the corresponding  eigenvalues $\nu_\alpha <1$  (or   $ \prod_{\nu_\alpha <1} \,  \nu_\alpha $ when there is more then one) 
as a function of the total gain $\gbar z$.  Finally, table  \ref{table_sympl4sq} lists the symplectric spectra for the four-mode square state \eqref{P4sq}. 
\begin{table}[ht]
\renewcommand{\arraystretch}{1.4}
\setlength{\tabcolsep}{5pt}
\begin{center}
\begin{tabular}{ | c | l @{\hspace{3pt}}  l  l | c | } 
  \hline
 Partition  & 
\multicolumn{3}{|c|}{Symplectic spectrum of $V^{\PTA  A} $}  &  $\nu_\alpha <1 $   \\ 
A & spectrum & \multicolumn{2} {l|}{eigenvalues}  & \\
  \hline
\hline 
$\{ 0 \} $   & $ \left( \begin{smallmatrix} e^{-2 \gbar z}, & 1, &e^{2 \gbar z}  \end{smallmatrix} \right)   $ &   
& 
&  $e^{-2 \gbar z }$ \; none \\ 
\hline
$\{ 1 \} $  
& $ \left( \begin{smallmatrix} \nu_-, & 1,   &\nu_+ \end{smallmatrix} \right)  $    
 & 
  $ \nu_{\pm} = b \pm \sqrt{b^2-1}   $
& \footnotesize{ $\bullet$ $ b=1 + 2  [\cos\theta \sinh {( \gbar z)}]^2 $}
 & $\nu_-$ \; $g_1 \ne 0$   \\ 
\hline
$\{ 2 \} $  
& $ \left( \begin{smallmatrix} \nu_-, & 1,   &\nu_+ \end{smallmatrix} \right)  $    
 & 
  $ \nu_{\pm} = b \pm \sqrt{b^2-1}   $
& \footnotesize{ $\bullet$ $ b=1 + 2  [\sin \theta \sinh {( \gbar z)}]^2 $}
 & $\nu_-$   \; $g_2 \ne 0$ \\ 
  \hline
\end{tabular}
\end{center}
\caption{  {\bf Three-mode coupling}\eqref{3state}: symplectic spectrum of the partial transpose matrix. The first column indicates the partition, via the Alice set A;  the second column describes the symplectic spectrum of $V^{\PTA A }$,  where     $\theta $  and $\gbar $ are the parameters of the Bloch-Messiah decomposition  in Eqs.\eqref{theta} and \eqref{gbar}. The last column gives the smallest  eigenvalue   and the conditions for $\nu_{\alpha} <1$ ensuring inseparability.
  }
\label{table_sympl3}
\end{table}

\begin{table}[h]
\renewcommand{\arraystretch}{1.4}
\setlength{\tabcolsep}{3pt}
\begin{center}
\begin{tabular}{ | c | l @{\hspace{8pt}}   l  | c  c| } 
  \hline
 Partition  & 
\multicolumn{2}{c}{Symplectic spectrum of $V^{\PTA  A} $}  & \multicolumn{2} {|c|}{ $\nu_\alpha  \le 1 $}   \\ 
A & Spectrum & Eigenvalues & & \\
  \hline
\hline 
$\{ 1 \} $ or $  \{ 2 \} $  & $ \left( \begin{smallmatrix} \nu_-, & 1, & 1,  &\nu_+ \end{smallmatrix} \right)   $ &   
$  \nu_{\pm} = \sqrt{ b \pm \sqrt{b^2-1}} $  
\quad \footnotesize{ $\bullet$ $ b=2 \left[\cosh (2 \rS) \cos^2{\gamma} + \cosh (2\rD) \sin^2 \gamma \right]^2 -1  $} 
&  $\nu_-$ & none\\ 
\hline
$\{ 3 \} $ or $ \{ 4 \} $
& $ \left( \begin{smallmatrix} \nu_-, & 1, & 1,  &\nu_+ \end{smallmatrix} \right)  $    
 & 
  $ \nu_{\pm} = \sqrt{b \pm \sqrt{b^2-1} }  $
\quad  \footnotesize{ $\bullet$ $ b=2 \left[\cosh (2 \rS) \sin^2{\gamma} + \cosh (2\rD) \cos^2 \gamma \right]^2 -1  $} 
 & $\nu_-$   &  $g_2 \ne 0$\\ 
\hline
 $\{ 1, 2\} $
 &  $ \left( \begin{smallmatrix} \nu_-, &  \nu_-,   &   \nu_+,  &\nu_+ \end{smallmatrix} \right)  $  
 & $ \nu_\pm =  \sqrt{1 + a^2} \pm a $ 
 \quad  \footnotesize{ $\bullet$  $a=  \sin(2\gamma) \sinh (\rD + \rS)  $}   
& \parbox[c][0.75cm]{0.7cm}{$\nu_- $ $  \nu_-$} 
&  $g_2 \ne 0$ \\
\hline
 $\{ 1, 3 \} $
& \multicolumn{2}{ l |} {$  ( \begin{smallmatrix} e^{-2\rS}, & e^{-2\rD}, & e^{2\rD},  &  e^{2\rS}\end{smallmatrix}) $ } 
  & \parbox[c] [0.8cm]{0.7cm}{ $ e^{-2\rS} $ $  e^{-2\rD} $ } & none \\
\hline
\multirow{2} {*}  { $\{ 1, 4 \}  $ } 
& \multirow{2} {*}   { $ \left( \begin{smallmatrix} \nu_-, &  \nu_+,   &   \mu_-,  &\mu_+ \end{smallmatrix} \right)  $ }
 &  $\nu_\pm = e^{-c} \sqrt{p \pm \sqrt{p^2-1}} $ 
\quad \quad  \, \footnotesize{ $\bullet$ $ c= \rS-\rD \ge 0  $}   & \multirow{2}{*}   {$\nu_-$}  &  \multirow{2}{*} { $g_1 \ne 0$}    \\ 
&  &$\mu_\pm = \sqrt{ p  e^{2c} \pm  e^{-2c} \sqrt{p^2-1}} $ 
 \;  \footnotesize{ $\bullet$ $ p=1 + 2 [\cos (2\gamma)  \sinh(\rD+\rS) ]^2   $}  &
&   \\
  \hline
\end{tabular}
\end{center}
\caption{ {\bf Four-mode linear coupling}\eqref{P4A}:  symplectic spectrum of the partial transpose. The first column indicates the bipartition, via the Alice set A;  the second column describes the symplectic spectrum of $V^{\PTA A }$,  where  $\gamma $  and  $r_{\SD} = \LSD z$ are the parameters of  the Bloch-Messiah decomposition of the state  in  Eq.\eqref{LSD}. The last column lists the  eigenvalues $\nu_{\alpha}  \le 1$,  and the conditions that guarantee inseparability through  $\nu_{\alpha} <1$. }
\label{table_sympl4}
\end{table}

\begin{table}
\renewcommand{\arraystretch}{1.4}
\setlength{\tabcolsep}{3pt}
\begin{center}
\begin{tabular}{ | c | l @{\hspace{8pt}}   l  | c  c| } 
  \hline
 Partition  & 
\multicolumn{2}{c}{Symplectic spectrum of $V^{\PTA  A} $}  & \multicolumn{2} {|c|}{ $\nu_\alpha  < 1 $}   \\ 
A & Spectrum & Eigenvalues & & \\
  \hline
\hline 
$\{ j =1,.. 4 \} $   & $ \left( \begin{smallmatrix} \nu_-, & 1, & 1,  &\nu_+ \end{smallmatrix} \right)   $ &   
$  \nu_{\pm} = \sqrt{ b \pm \sqrt{b^2-1}} $  
\quad \footnotesize{ $\bullet$ $ b= \half \left[\cosh (2 \rS)  + \cosh (2\rD) \right]^2 -1  $} 
&  $\nu_-$ & none\\ 
\hline
 $\{ 1, 2\} $
 &  $ \left( \begin{smallmatrix} \nu_-, &  \nu_-,   &   \nu_+,  &\nu_+ \end{smallmatrix} \right)  $  
 & $ \nu_\pm = e^{ \pm  \left| \rS+ \rD\right|}$    \quad \footnotesize{$\bullet$ $ |\rS + \rD| = 2 |g_1 + g_2| z$}
& \parbox[c][0.75cm]{0.7cm}{$\nu_- $ $  \nu_-$} 
&  $g_1 + g_2 \ne 0$ \\
\hline
 $\{ 1, 3 \} $
& \multicolumn{2}{ l |} {$  ( \begin{smallmatrix} e^{-2\rS}, & e^{-2|\rD|}, & e^{2|\rD|},  &  e^{2\rS}\end{smallmatrix}) $ } 
  & \parbox[c] [0.8cm]{0.7cm}{ $ e^{-2\rS} $ $  e^{-2 |\rD|} $ } & none \\
\hline
 $\{ 1, 4\} $
 &  $ \left( \begin{smallmatrix} \nu_-, &  \nu_-,   &   \nu_+,  &\nu_+ \end{smallmatrix} \right)  $  
 & $ \nu_\pm = e^{ \pm  ( \rS- \rD) }$   \quad \footnotesize{$\bullet$ $ \rS- \rD =|g_1|z + |g_2| z$}
& \parbox[c][0.75cm]{0.7cm}{$\nu_- $ $  \nu_-$} 
&  none\\
\hline
\end{tabular}
\end{center}
\caption{ {\bf Four-mode square coupling}\eqref{P4sq}: symplectic spectrum of the partial transpose. The first column gives the partition via  the Alice set A;  the second column describes the symplectic spectrum of $V^{\PTA A }$,  where    $r_{\SD} = \LSD z$ are the squeeze parameters  in  Eq.\eqref{LSDsq}. The last column lists the  eigenvalues $\nu_{\alpha}  \le 1$,  and the conditions that guarantee inseparability through  $\nu_{\alpha} <1$. }
\label{table_sympl4sq}
\end{table}

\newpage
\section{Heisenberg-like bounds for the 4-mode linear state}
\label{app_bounds}
Table \ref{table_bound4} reports our explicit calculations  for the violation  of the  Heisenberg-like inequalities \eqref{bound4} for the  variances of the nonlocal observables \eqref{eta4},  which provide sufficient criteria for the entanglement of the 4-mode linear coupling described by Eq. \eqref{P4A}.  For each partition, we  calculated the four commutators at the r.h.s of the inequalities\eqref{bound4}: these are reported in the central column of table \ref{table_bound4}. By using the explicit expressions of the variances reported in Eq. \eqref{vareta4},    we chose among the four bounds  the best one, i.e. the one which is violated  for the smallest value  of  $\gbar z$ (highlighted in gray in the table): the last two columns show then the corresponding sufficient condition for separability.  This has  been explicitly written 
 as a function of the squeeze parameters  $\rS= \LS z$ and  $\rD =\LD z$, by substituting  
$\cos^2 \gamma= \frac{\rS}{\rS + \rD}$ and  
$\sin^2 \gamma= \frac{\rD}{\rS + \rD}$ [see Eq. \eqref{LSD}] .  Finally, the explicit expression of the squeeze parameters in Eq.\eqref{LSD} was used to calculate the curves reported in  Fig.\ref{fig_bounds4}. 
\begin{table}
\renewcommand{\arraystretch}{1.8}
\setlength{\tabcolsep}{4pt}
\begin{center}
\begin{tabular}{ | c | c @{\hspace{1pt}}  |c|  c | c  c| } 
  \hline
 Partition  & 
\multicolumn{3}{|c|}{$B_{\alpha \beta}=  \left| \left[ \hat \eta ( \PT_A \vd_\alpha) , \hat  \eta (\PT_A \vd_\beta^\prime ) \right] \right| 
  $  }                                  &  Inseparability  & Needs \\ 
A & $B_{\sigma \sigma}$  &  $B_{\sigma \delta } = B_{\delta \sigma}$  &  $B_{\delta \delta}$   & & \\
 \hline
$\{ 1 \} $ or $  \{ 2 \} $   
& \cellcolor[gray]{0.9} $ 2 \cos^2 \gamma $ 
&$  2 \cos  \gamma  \sin  \gamma   $  
&  $ 2 \sin^2  \gamma $ 
&   \parbox[c] [0.5cm]{3cm}{  $ e^{-2 \rS} < \frac{ \rS}{\rS + \rD} $} & none\\ 
\hline
$\{ 3 \} $ or $ \{ 4 \} $
& $ 2 \sin^2  \gamma  $ 
&$  2 \cos  \gamma  \sin  \gamma   $  
& \cellcolor[gray]{0.9}   $ 2 \cos^2  \gamma $ 
 &  \parbox[c] [0.5cm]{3cm}{ $ e^{-2 \rD} < \frac{ \rS}{\rS + \rD} $} & $g_2 \ne 0$\\ 
\hline
 $\{ 1, 2\} $
& $ 0 $ 
&  \cellcolor[gray] {0.9} $   4 \cos  \gamma  \sin  \gamma   $  
&  $ 0$ 
&  \parbox[c] [0.5cm]{3cm}{ $ e^{- ( \rS + \rD) } < \frac{2 \sqrt{ \rS \rD} }{\rS + \rD} $ } 
&  $g_2 \ne 0$ \\
\hline
 $\{ 1, 3 \} $
& \cellcolor[gray] {0.9}  $  2  $ 
&$  0 $  
&  $ 2 $ 
& \parbox[c] {1.5cm}{ $ e^{-2\rS} <  1 $  } & none \\
\hline
 $\{ 1, 4\} $
& \cellcolor[gray] {0.9}   $  2 (\cos^2 \gamma  -\sin^2  \gamma )  $ 
&$ 0  $  
&  $  2 (\cos^2  \gamma  -\sin^2  \gamma )  $ 
&  \parbox[c] [0.5cm]{3cm}{  $ e^{-2 \rS} < \frac{ \rS -\rD}{\rS + \rD} $}
&  $g_1 \ne 0$ \\
  \hline
\end{tabular}
\end{center}
\caption{  {\bf Four-mode linear coupling}\eqref{P4A}: sufficient criteria for inseparability  based on the violation of the Heisenberg-like bounds \eqref{bound4}.  The first column indicates the bipartition, via the Alice set A;  the second column gives the 4 bounds, where we highlighted the one which is violated for smaller values of the gain $\gbar z$.  The last column lists the  corresponding conditions sufficient for  inseparability. $\gamma$  and  $r_{\SD} = \LSD z$ are the parameters of  the Bloch-Messiah decomposition in \eqref{LSD}.  }
\label{table_bound4}
\end{table}


%
\end{document}